\documentclass[aps]{revtex4}
\begin{document}

\title{
      Nuclear fission with mean-field instantons
         }

\author{Janusz Skalski }

\affiliation{
 So\l tan Institute for Nuclear Studies,\\
ul. Ho\.za 69, PL - 00 681, Warsaw, Poland \\
 e-mail: jskalski@fuw.edu.pl, tel/fax: (48 22) 621 60 85
 }

\date{ \today}

\begin{abstract}
We present a description of nuclear spontaneous fission, and generally of quantum tunneling, in
terms of instantons - periodic imaginary-time solutions to time-dependent mean-field equations -
that allows for a comparison with more familiar and used generator coordinate (GCM) and adiabatic
time-dependent Hartree-Fock (ATDHF) methods. It is shown that the action functional whose value
for the instanton is the quasiclassical estimate of the decay exponent fulfils the minimum principle
when additional constraints are imposed on trial fission paths.
 In analogy with mechanics, these are conditions of energy
conservation and the velocity-momentum relations. In the adiabatic limit
 the instanton method reduces to the time-odd ATDHF equation,
 with collective mass including the time-odd Thouless-Valatin term,
 while the GCM mass completely ignores
  velocity-momentum relations. This
implies that GCM inertia generally overestimates instanton-related decay rate.
 The very existence of the minimum
principle offers a hope for a variational search for instantons.
After the inclusion of pairing, the
instanton equations and the variational principle can be expressed
in terms of the imaginary-time-dependent Hartree-Fock-Bogolyubov
(TDHFB) theory. The
 adiabatic limit of this theory reproduces ATDHFB inertia.

\end{abstract}

\pacs{PACS number(s):  }

\maketitle

\section{Introduction}
Decay of a metastable state of a system of interacting fermions or bosons is an important phenomenon relevant to
nuclear, atomic and condensed matter physics. The calculation of decay rate requires the exact knowledge of the wave
function in the proper asymptotic region which is usually very difficult to achieve for many-body systems. In fact,
very often the only feasible description of systems including hundreds or more
 particles relies on the quantum mean-field
theory. Unfortunately, such theory does not contain quantum tunneling. This gives rise to a notorious arbitrariness
in calculations of decay rates or half-lives which concerns a selection of
 relevant degrees of freedom and prescriptions for potential
 and inertia parameters.

 Specifically, within the Hartree-Fock (HF) method, static equations give only
 saddle points on energy landscape
   \begin{equation}
   \label{mfh1}
 {\cal H}[\psi^*,\psi]=\int dx \sum_k
 \frac{\hbar^2}{2m}\nabla \psi_k^*\nabla \psi_k +
 {\cal V}[\psi^*,\psi] ,
 \end{equation}
 with ${\cal V}[\psi^*,\psi]$ being potential energy, so one has to resort to
 the time-dependent HF (TDHF) equations for dynamics
\begin{equation}
\label{eqsp1}
 i \hbar \partial_t \psi_k(t) =
  {\hat h}(t)\psi_k(t) =
- \frac{\hbar^2}{2m} \nabla^2 \psi_k(t) + \frac{\delta {\cal V}}{\delta
  \psi^*_k(t)} ,
\end{equation}
 with the mean-field single-particle (s.p.) Hamiltonian ${\hat h}(t)$ given by
  ${\hat h}[\psi^*(t),\psi(t)]\psi_k(t)=\delta{\cal H}/\delta\psi^*_k(t)$,
 and the self-consistent s.p. potential ${\hat V}(t)$ given by
 ${\delta {\cal V}}/{\delta \psi^*_k(t)}={\hat V}(t)\psi_k(t)$.
 For the case of energy ${\cal H}$ given by a density functional we assume 
  in the following (if not indicated otherwise) that it has properties of the 
 expectation value of the Hamiltonian.
   Although Eqs.(\ref{eqsp1}) look like the Schr\"odinger equations,
 in fact, they are classical field equations, due to a nonlinear
 dependence of ${\hat h}$ on $\psi_k$.
 Energy (\ref{mfh1}) and overlaps $\langle \psi_k\mid \psi_l\rangle$ are
 conserved by Eqs.(\ref{eqsp1}).
 The former forbids a tunneling within TDHF, i.e. an escape from a minimum
  of ${\cal H}$ with energy lower than the saddle. Evidently, this comes about
 by projection of the full many-body theory onto Slater states.

 A quasiclassical treatment of quantum tunneling within the many-body
 mean-field theory, which is a natural generalization of the Gamow treatment
  of alpha-decay to an infinite-dimensional system of fields, leads to
  instantons - periodic imaginary-time solutions to TDHF equations \cite{LNP}.
 This method exploits an idea of trajectories evolving in imaginary time
 \cite{Col} which emerge from the stationary-phase approximation to the
 path-integral expression for $Tr(E-{\hat H})^{-1}$. The decay rate of
 a metastable state is proportional to $\exp(-S/\hbar)$, where $S$ is
 action for the optimal instanton. We do not consider here a prefactor
 coming from quantum fluctuations around the optimal path.

  For a particle in an external potential such optimal decay trajectory
 describes classical motion in the inverted potential. It starts at the
    metastable state (being a local maximum of the inverted well)
   and returns there after bouncing from the inverted barrier;
    hence the name `bounce'.
  For a system of interacting fermions, one has to
 transform TDHF Eqs.(\ref{eqsp1}) to imaginary time, i.e., formally,
 $t\rightarrow -i\tau$.
 Under this transformation, $\psi\rightarrow \psi(x,-i\tau)=\phi(x,\tau)$
 and $\psi^*\rightarrow \psi(x,-i\tau)^* = \phi(x,-\tau)^*$ \cite{LNP,JN}.
 It follows that density $\rho(x,t)=\psi^*(x,t)\psi(x,t)$ transforms to
 $\rho(x,\tau)=\phi(x,-\tau)^*\phi(x,\tau)$. This has important consequences.
 First, the mean-field equations in imaginary time \cite{LNP,JN,super}:
\begin{equation}
\label{eqsp}
 \hbar\frac{\partial\phi_k}{\partial\tau}(\tau) =
 -\left({\hat h}(\tau)-\epsilon_k\right)\phi_k(\tau)=
\frac{\hbar^2}{2m} \nabla^2 \phi_k(\tau) -
 \frac{\delta {\cal V}}{\delta \phi_k^*(-\tau)}+\epsilon_k\phi_k(\tau) ,
\end{equation}
 are non-local in $\tau$, as ${\cal V}$ as well as ${\hat h}(\tau)={\hat h}
 [\phi^*(-\tau),\phi(\tau)]$ depend on both $\phi(x,\tau)$ and
 $\phi(x,-\tau)$. Second, density $\rho(x,\tau)$, generally complex or
   piecewise negative, does not correspond to any Slater determinant,
  unlike in the real-time dynamics.
In analogy with TDHF, Eqs.(\ref{eqsp}) conserve energy
 ${\cal H}(\tau)={\cal H}[\phi^*(-\tau),\phi(\tau)]$, with:
   \begin{equation}
   \label{mfh}
 {\cal H}(\tau)=\int dx \sum_k
 \frac{\hbar^2}{2m}\nabla \phi_k^*(-\tau)\nabla \phi_k(\tau) +
 {\cal V}[\phi^*(-\tau),\phi(\tau)] .
 \end{equation}
 The above formula means that one obtains ${\cal H}(\tau)$ replacing everywhere
 $\psi^*_k(t)$ by $\phi^*_k(-\tau)$ in the usual form of the
 energy functional.
 Since the Hamiltonian is hermitean, ${\hat H}^+={\hat H}$, it follows
 that ${\cal H}(-\tau)={\cal H}^*(\tau)$
 and the mean-field Hamiltonian ${\hat h}(\tau)$, defined by ${\hat h}(\tau)
 \phi(\tau)=\delta{\cal H}/\delta\phi^*(-\tau)$,
 fulfils the condition ${\hat h}(-\tau)={\hat h}^+(\tau)$. The latter
 ensures that Eqs. (\ref{eqsp}) {\it without
 the $\epsilon_k\phi_k$ term} conserve the overlaps:
$\frac{d}{d\tau}\langle \phi_i(-\tau)\mid\phi_j(\tau)\rangle=0$.
 The complete Eqs.(\ref{eqsp}) still conserve diagonal overlaps, while giving
 the exponential time-dependence to the off-diagonal ones. However, those
 overlaps remain zero for all $\tau$, if equal zero at some $\tau$.

 As usual, the saddle point approximation to the path integral leads to
  the periodicity condition for the optimal trajectories. Hence,
  bounce is a periodic intanton,
\begin{equation}
  \phi_k(T/2) = \phi_k(-T/2)  ,
 \end{equation}
 and the periodicity is enforced by the $\epsilon_k\phi_k$ term in the
 Eqs.(\ref{eqsp}).
   The physical context imposes the specific boundary conditions on bounce.
  For a description of the decay of a metastable ground state,
 the initial (and thus also the final) states have to be chosen equal to
 the HF solutions $\psi^{HF}_k$ at the metastable minimum,
 $\phi_k(T/2)=\phi_k(-T/2)=\psi^{HF}_k$, with total energy $E_{gs}$, and
 the parameters $\epsilon_k$ must be equal to the HF s.p.
 energies at this minimum.
 The s.p. states $\phi_k(\tau=0)$ form some normal (as $\phi_k^*(-\tau)=
 \phi_k^*(\tau)$ at $\tau=0$) HF state at energy ${\cal H}=E_{gs}$ on
 "the other side of the barrier".
 The periodicity condition together with the initial condition fix the
 particular constant values of the overlaps:
 \begin{equation}
 \label{overdel}
 \langle \phi_i(-\tau)\mid\phi_j(\tau)\rangle=\delta_{ij} .
 \end{equation}

  Decay exponent is given by \cite{LNP,JN}:
   \begin{equation}
   \label{S}
 S =
  \hbar\int_{-T/2}^{T/2} d\tau \sum_k \langle\phi_k(-\tau)\mid
 \frac{\partial \phi_k}{\partial \tau}(\tau)\rangle .
 \end{equation}
  Bounce penetrates the static barrier, impermeable for real-time
  solutions at the same energy, practically in a finite time interval
  around $\tau=0$ and becomes infinitely slow close to the endpoints,
  so that $T$ extends to infinity \cite{LNP,super,JN}.
 Eqs.(\ref{eqsp}) determine both decay channels
 and decay probabilities.  No additional assumptions are necessary,
 as they form a complete quasiclassical solution to the tunneling problem
 within the mean-field theory.

 Up to now, solutions of Eqs.(\ref{eqsp}) have been obtained only for
 relatively simple systems \cite{LNP,Arve,PN,JN,super,JS}.
 The task of finding instantons seems
 rather hopeless without a special treatment: to handle non-locality in $\tau$
 one could try to solve Eqs.(\ref{eqsp}) together with
 \begin{equation}
  \label{eqspant}
 -\hbar \frac{\partial[\phi_k(-\tau)]}{\partial\tau} +
  \left({\hat h}(-\tau)-\epsilon_k\right)\phi_k(-\tau) = 0 ,
\end{equation}
 describing instantons evolving backwards, obtained from (\ref{eqsp}) by using
 the identity $(\partial_{\tau}f)(-\tau)=-\partial_{\tau}(f(-\tau))$.
 However, Eqs.(\ref{eqspant}) describe the inverse diffusion
 [cf signs of time and spatial derivatives in (\ref{eqsp}) and
 (\ref{eqspant})], which leaves no hope for a stable solution.
 The problem seems more difficult than a search for periodic solutions of
 the real-time TDHF equations, which is known to be difficult enough.
 In the presented form, instanton approach did not lead to any comparisons
 with the existent studies of fission, which are mostly based either on
 the generator coordinate method (GCM) in the gaussian overlap approximation
 (GOA) or on the adiabatic TDHF (ATDHF) method, either in its extreme cranking
 or some more refined version.

 In this work we present the instanton method in familiar terms of the HF or
 HFB theory. This helps to grasp similarities and differences between this
 and other methods and to clarify their interrelations. In particular, the
 instanton turns out to be an analogue of the self-consistent TDHF in the
 representation of the time-even and time-odd components of the density
 matrix \cite{BV78}. Both produce the same inertia in the ATDHF limit,
 when one includes time-odd components only to the first order.

 Moreover, it turns out that instanton action Eq.(\ref{S}) is the minimum value
 of the action functional over properly constrained set of trial fission
 trajectories defined in the space of Slater determinants \cite{JS1}. Thus, any fission
 path which satisfies these constraints provides the upper bound for the
 decay exponent. This offers a hope for a variational
approach to finding instantons. One may also expect that a good estimate of
 action may be easier to find than that of the instanton itself.

 The starting point is the realization that Eqs.(\ref{eqsp}) describe two
 different sets of Slater determinants, bra $\Phi(-\tau)$ built out of
 $\phi_k(-\tau)$, and ket $\Phi(\tau)$ built out of
 ${\phi_k(\tau)}$, while energy ${\cal H}$ is equal to the off-diagonal energy
 overlap kernel in the sense of GCM,
 $\langle\Phi(-\tau)\mid {\hat H} \mid \Phi(\tau)\rangle/\langle\Phi(-\tau)
 \mid\Phi(\tau)\rangle$ \cite{JS1}.
 It is the difference between bra and ket that makes
 barrier tunneling possible and allows for the conservation of energy
 Eq.(\ref{mfh}). The energy overlap kernel reduces to
 $\langle\Phi(-\tau)\mid {\hat H} \mid \Phi(\tau)\rangle$
 owing to the choice of the overlap value
 $\langle\Phi(-\tau)\mid\Phi(\tau)\rangle=1$ that follows from
 $\langle \phi_k(-\tau)\mid\phi_l(\tau)\rangle=\delta_{kl}$.
 However, the overlap of the normalized bra and ket,
 $(\langle\Phi(-\tau)\mid\Phi(-\tau)\rangle
 \langle\Phi(\tau)\mid\Phi(\tau)\rangle)^{-1/2}$, is smaller than 1.
 Bounce may be thought of as one of many trial tunneling paths $\{\phi_k\}$,
  each given as two sets of wave functions,
 $\{\phi_{1 k}(\tau)\}$ and $\{\phi_{2 k}(\tau)\}$, defined
 on the interval $[0,T/2]$, and related to the variables of Eqs.(\ref{eqsp}):
  \begin{equation}
  \label{path}
   \phi_k(\tau) = \left\{ \begin{array}{cc}
            \phi_{1 k}(-\tau) & \mbox{for} \; \tau<0 , \\
            \phi_{2 k}(\tau)   & \mbox{for} \; \tau>0   \\
                   \end{array} \right\} .
  \end{equation}
 At $\tau=0$ both $\Phi_1$ and $\Phi_2$ are equal to some
  constrained HF (CHF) state $\Phi(0)$ at the outer slope of the barrier
 with the constraint $-\partial_{\tau}\Phi(0)$. The Eqs.(\ref{eqsp}) and
  (\ref{eqspant}),
 rewritten in terms of $\phi_{1 k}(\tau)$ and $\phi_{2 k}(\tau)$ are
\begin{eqnarray}
 \label{eqpa}
 \hbar\partial_{\tau}\phi_{2 k}+({\hat h}(\tau)-\epsilon_k)\phi_{2 k} &=& 0 ,
 \\ \nonumber
 -\hbar\partial_{\tau}\phi_{1 k}+({\hat h}^+(\tau)-\epsilon_k)\phi_{1 k} &=&0 ,
 \\ \nonumber
\end{eqnarray}
 with ${\hat h}(\tau)={\hat h}[\phi_{1 k},\phi_{2 k}]$. It should be clear
 that one can equally well use fields restricted to $[-T/2,0]$.

 The paper is organized as follows: The main results for the HF instanton 
 method are contained in sections IV-VII. These are: the variational principle
 (sect. IV), the formulation in terms of coordinates and momenta
 and comparison to the cranking method (sect. V), the introduction of 
 special variables in the form of the time-even density matrix 
 and the time-odd hermitean operator that make plain the adiabatic limit of the 
 theory (sect. VI) and the demonstration that the GCM+GOA action follows 
  from that for instanton after neglecting the velocity-momentum relations
 (sect. VII).  
  In section VIII these results are generalized to systems with pairing. 
 Section III prepares useful formulas for later sections. 
  Section II introduces some unusual features of the instanton method.
 Conclusions are given in section IX.

 \section{General overview}

  A few comments on several unusual features of the instanton equations may be
  helpful.

 As the linear combination of the s.p. wave functions changes their Slater
 determinant only up to a factor, one may expect that the instanton equations
 may be more general than Eqs.(\ref{eqsp}) which fix in a specific way
 lengths and angles among each of the sets $\{\phi_{1 k}\}$ and
 $\{\phi_{2 k}\}$ separately.
 Such a more general equation will imply a more general expression for
 the instanton action than Eq.(\ref{S}), and both will be given in the
  next section.

 The non-local in time form of the instanton equations follows directly from
 the transformation of the standard variational principle of the TDHF theory,
$\delta\int\langle\Psi(t)\mid i\hbar\partial_t-{\hat H}\mid\Psi(t)\rangle dt=0$,
 to imaginary time,
$\delta\int\langle\Phi(-\tau)\mid\hbar\partial_{\tau}+{\hat H}\mid\Phi(\tau)
 \rangle d\tau=0$. The Eqs.(\ref{eqsp}) and (\ref{eqspant}),
 without the periodicity-fixing terms, have the canonical form in strange
 variables
 \begin{eqnarray}
 \hbar\frac{\partial\phi_k(\tau)}{\partial{\tau}}&=& -\frac{\delta{\cal H}}
 {\delta\phi_k^*(-\tau)}, \\  \nonumber
 \hbar\frac{\partial[\phi_k^*(-\tau)]}{\partial{\tau}}&=& \frac{\delta{\cal H}}
 {\delta\phi_k(\tau)}, \\  \nonumber
 \end{eqnarray}
 none of which has a determined time parity. The usual canonical variables
 are $\tau$-even coordinates and $\tau$-odd momenta.
 Such standard coordinates and momenta may be introduced
 by a change of variables with the resulting equations of motion local in
 time and canonical in form (section V).
 It should be stressed that a local form of the instanton equations does not
 facilitate their solution, but makes easier their comparison to other
 theories of the large amplitude collective motion (LACM).
 One possibility is given by \cite{JN}:
 $\phi_k=\sqrt{\rho_k}\exp(-\chi_k)$, whith $\rho_k$ time-even
 and $\chi_k$ time-odd. For one real-valued wave function and potential
 energy being a functional of density, ${\cal V}[\rho]$, one obtains
 the continuity and ``fluid velocity'' equations, as for the density-phase
 representation of the Schr\"odinger equation.
 Energy ${\cal H}$ becomes:
\begin{equation}
 {\cal H} = \frac{\hbar^2}{m}\int dx\left[-\frac{\rho(\nabla\chi)^2}{2}+
 \frac{(\nabla\rho)^2}{8\rho}\right]+{\cal V}[\rho]  ,
\end{equation}
 where the minus sign shows the role of $\chi$ in the lowering of energy down
 to $E_{gs}$ in the barrier region. From the boundary conditions, symmetries
 of $\rho$ and $\chi$, and the continuity equation, one obtains action:
 $S=(\hbar^2/m)\int d\tau dx \rho(\nabla\chi)^2$.
 For simple systems, like the Bose-Einstein condensate of $^7$Li atoms,
 this framework allows for the exact treatment of the collapse of
 the metastable state \cite{JS}. However, the density-phase variables
 seem unsuitable for fermions due to the spinor structure and the
 rearrangement of nodes of s.p. wave functions along the barrier
 that makes phases singular. More appropriate variables are defined in
 sections V and VI.

 The other peculiarity of Eqs.(\ref{eqsp}) is that they may be thought of
 as describing
 a forced motion: The mean field ${\hat h}$ that causes the evolution of
 $\Phi(\tau)$ depends on $\Phi(-\tau)$ so one may say that one state drags
 the other. More specifically, as climbing the barrier is impossible without
 an external drive, the drag is necessary at the beginning of the motion
 from the metastable state through the barrier and at the beginning of the
 return motion to the metastable minimum.
 Action Eq.(\ref{S}) is given
 by the integral of the scalar product between the change in the driven state
 and the state that drives it. For motions for
 which the result of the dragging is fixed by the instanton boundary
 conditions there must be some minimal dragging
 that causes this (fixed) result. Hence, one can expect that there is a
 minimum principle which selects instantons. If
 so, then solving Eqs.(\ref{eqsp}) and finding decay exponent could
 be done by a minimization of a functional. The functional
 is practically given by Eq.(\ref{S}). What remains to be done is to learn the
 necessary additional constraints which make this action minimal for
 instantons.

  In fission studies, mean-field states are parametrized by expectation
 values of observables that provide coordinates along the barrier,
 called deformations.
 Consider as an example the quadrupole moment ${\hat Q}$.
 For bounce states $\Phi(\tau)$ one has
 two possible labels:
 Within the imaginary-time formalism, a natural choice is:
 $Q(\tau)=\langle\Phi(-\tau)\mid{\hat Q}\mid\Phi(\tau)\rangle=
 \sum_k\langle\phi_k(-\tau)\mid{\hat Q}\mid\phi_k(\tau)\rangle$.
 Since ${\hat Q}$ is hermitean,
 $Q(-\tau)=Q^*(\tau)$, and ${\dot Q}(\tau)=dQ/d\tau=
 \sum_k\langle\phi_k(-\tau)\mid
 [{\hat h}(\tau),{\hat Q}]\mid\phi_k(\tau)\rangle$, with ${\dot Q}(-\tau)=
 -{\dot Q}^*(\tau)$. Thus the real part of ${\dot Q}(\tau)$ fixes $\tau=0$
 as the return (or bounce) point.
  Another possibility is to trace deformation of the
  normalized state $\Phi(\tau)$, $q(\tau)=\langle\Phi(\tau)\mid{\hat Q}\mid
 \Phi(\tau)\rangle/\mid\Phi(\tau)\mid^2$. Generally, $q(\tau)\ne Q(\tau)$ 
 and $q(\tau)\ne q(-\tau)$, except for $\tau=0$ and $\pm T/2$. 

 Instanton cannot depend solely on a time-even variable like the real part of
 $Q$, as then Eq.(\ref{eqsp}) at $\tau=0$ would require a static HF solution
 without constraints which cannot exist on the barrier slope.
 One can observe that the derivative ${\dot q}(\tau=0)$ is equal to 
 $2\Re\sum_k\langle[\partial_{\tau}
 \phi_k(0)]_{\perp}\mid{\hat Q}\mid\phi_k(0)\rangle$, where
 $[\partial_{\tau}\phi_k(0)]_{\perp}$ is perpendicular to all $\phi_k(0)$.
 Considering $\Phi(0)$ as a stationary HF state with the
 constraint $-\partial_{\tau}\Phi(0)$, one can see that ${\dot
 q}(0)$, up to a positive constant, is the scalar product of two constraints:
  the one of instanton at $\tau=0$ Eqs.(\ref{eqsp}) and the other,
  $-{\hat Q}\Phi(0)$, the proper quadrupole constraint on the slope
  where $\partial {\cal H}/\partial Q<0$.
  Since $\phi_k(0)$ lives on this slope and has the quadrupole
  moment $Q(0)$, it must be close to some ${\hat Q}$-constrained HF state.
   Hence this scalar product and the derivative ${\dot q}(0)$ are very likely
  positive. Indeed, it was found positive in the simple model \cite{JS}.
  If so, the return point for the coordinate $q(\tau)$ is at 
  $\tau>0$, which means that at $\tau=0$ the quadrupole moment of the 
  normalized state $\Phi_2$ still increases while that of $\Phi_1$ 
 (istanton evolving backwards) decreases.
  Moreover, as show calculations for simple systems,
 states $\phi_{1k}$ and $\phi_{2k}$ with the same $q$ are different.
  Thus, neither $Q$, nor $q$ are sufficient as labels for bounce.

 In general, the instanton mean field is not hermitean. The condition it
 satisfies, ${\hat h}(-\tau)={\hat h}^+(\tau)$, imposes the
 following conditions on the hermitean and antihermitean parts of its standard
 decomposition ${\hat h}(\tau)={\hat h}_R(\tau)+{\hat h}_A(\tau)$:
 ${\hat h}_R(-\tau)={\hat h}_R(\tau)={\hat h}_R^+(\tau)$, and
 ${\hat h}_A(-\tau)=-{\hat h}_A(\tau)={\hat h}_A^+(\tau)$.
 The antihermitean mean field ${\hat h}_A$ comes from $\tau$-odd components of
 densities appearing in energy ${\cal H}$, either in the form of the
 expectation value of ${\hat H}$ or in the form of energy functional.
  In the latter case, as for the Skyrme energy functional, the generic 
  contribution to ${\hat h}_A$ in the tunneling problem comes from the current 
  density ${\bf j}$. In the imaginary-time formalism, it takes a form:
 ${\bf j}(\tau)=\sum_k(\phi_k(\tau)\nabla\phi^*_k(-\tau)-
 \phi_k^*(-\tau)\nabla\phi_k(\tau))/2$, which follows from this part of 
 Eqs.(\ref{eqsp}) that shows the continuity of the probability flow. 
  It follows that ${\bf j}(-\tau)=-{\bf j}^*(\tau)$.
 This differs by the factor $(-i)$ from the conventional
 current in the real-time TDHF. As a result,
 the time-odd contribution to the TDHF mean field $i\cdot{\bf j}\cdot\nabla$
 becomes $-{\bf j}\cdot\nabla$ in the imaginary-time formalism.
 Its antihermitean part is proportional to {\it the real part} of
 ${\bf j}(\tau)$, and the latter appears as soon
 as the real parts of functions $\phi_k(\tau)$ and $\phi_k(-\tau)$
 become different. 
 The time-odd mean field ${\hat h}_A$ is the immediate imaginary-time analogue 
 of the Thouless-Valatin potential in TDHF \cite{TV}, and we will use this
 name for it.


\section{Various forms of instanton action and equations}

  The value of $S$ which determines the fission probability
  relies only on a part of information contained in the bounce solution.
 By using general identities: $(\partial_{\tau}f)(-\tau)=-\partial_{\tau}
(f(-\tau))$, $\int_{-a}^a d\tau[f(\tau)-f(-\tau)]=0$, and the
 constancy of diagonal overlaps Eq.(\ref{overdel}) one can recast
 Eq.~(\ref{S}) into the following forms:
   \begin{eqnarray}
   \label{SR}
 S/\hbar & = & -\int_{-T/2}^{T/2} d\tau \sum_k
  \langle\phi_k(\tau)\mid\partial_{\tau}[\phi_k(-\tau)]\rangle \\ \nonumber
  & = & \Re \int_{-T/2}^{T/2} d\tau \sum_k
  \langle \phi_k(-\tau)\mid\partial_{\tau} \phi_k(\tau)\rangle  \\ \nonumber
  & = & 2\Re \int_{0}^{T/2} d\tau \sum_k
  \langle \phi_k(-\tau)\mid\partial_{\tau} \phi_k(\tau)\rangle . \\ \nonumber
 \end{eqnarray}
 The first equality shows that action for instanton evolving backwards in
 time, $\phi_k(-\tau)$, equals to minus action for the instanton.
 The second equality shows that instanton
 action is a real number; the third one expresses action in terms of variables
 $\phi_{1 k}$ and $\phi_{2 k}$ defined by Eqs.(\ref{eqpa}).

  Since $\partial_{\tau}\mid \phi_k\rangle =
 (\partial_{\tau}\ln\mid\phi_k\mid)\mid\phi_k\rangle+v$, with
  $v\perp\mid\phi_k(\tau)\rangle$, and $\phi_l(\tau)$ for all $l\ne k$ are
  perpendicular to $\phi_k(-\tau)$,
  the integrand $\langle\phi_k(-\tau)\mid\partial_{\tau}\phi_k(\tau)\rangle$
  is the sum of the full derivative plus the contribution from the
 component $[\partial_{\tau}\phi_k]_{\perp}$ of the
 derivative $\partial_{\tau}\phi_k$ orthogonal to the subspace spanned by
  all vectors $\left\{ \phi_k(\tau)\right\}_{k=1}^N$.
   After integration from $-T/2$ to $T/2$, only the latter contribution
 is left
   \begin{equation}
   \label{Sa}
 S =
  \hbar\int_{-T/2}^{T/2} d\tau \sum_k \langle [\phi_k(-\tau)]_{\perp}\mid
 [\frac{\partial \phi_k}{\partial \tau}(\tau)]_{\perp}\rangle ,
 \end{equation}
 where $[\phi_k(-\tau)]_{\perp}$ is the component of $\phi_k(-\tau)$
 perpendicular to $\left\{ \phi_k(\tau)\right\}_{k=1}^N$. This shows that
 $[\phi_k(-\tau)]_{\perp}$ are the essential variables conjugate to
 $\phi_l(\tau)$, while the components of $\phi_k(-\tau)$ in  the subspace
  $\left\{ \phi_k(\tau)\right\}_{k=1}^N$ are completely fixed by the
 overlap constraints Eq.(\ref{overdel}).

 As bounce $\Phi(\tau)$ is a closed cycle in the Hilbert space
 ($\Phi(-\tau)\neq\Phi(\tau)$ unlike for a line segment),
  action $S$ may be written in a form of the contour integral:
 \begin{equation}
 \label{Sfun}
  S= \hbar \oint \sum_k \langle \phi_k(-\tau)\mid d \phi_k(\tau)\rangle  ,
 \end{equation}
 which manifests reparametrization invariance of $S$: it does not depend
 at all on the instanton "speed". As can be seen from Eq.(\ref{Sfun}), the only
  important features are: the path traced by $\mid \phi_k\rangle$ in the vector
  space of s.p. states
 and the rule which associates pairs $\langle \phi_k(-\tau)\mid $
 and $\mid \phi_k(\tau)\rangle$. Reparametrizations of imaginary time,
 $\tau\rightarrow \theta(\tau)$, that are both invertible ($d\tau/d\theta>0$)
 and consistent with the association rule: $\tau(-\theta)=-\tau(\theta)$,
 ($\tau(-\Theta/2)=-T/2$, $\tau(\Theta/2)=T/2$), leave $S$ invariant.
 However, the reparametrized bounce, $\phi_k(\theta)$ is not a solution to
 Eqs.(\ref{eqsp}). Instead, it solves:
\begin{equation}
 \label{eqrep}
 \hbar \frac{\partial\phi_k}{\partial \theta}(\theta)
   + \left(\frac{d\tau}{d\theta}\right)({\hat h}(\theta)-\epsilon_k)
 \phi_k(\theta) = 0 .
\end{equation}

 One can recover action if bounce is known up to an $\tau$-dependent invertible
 linear transformation. Consider states $\psi_k(\tau)$ related to bounce
 $\phi_k(\tau)$ by means of such a transformation $N(\tau)$:
 \begin{equation}
 \label{linear}
  \phi_k(\tau)= \sum_l N_{l k}(\tau) \psi_l(\tau) .
 \end{equation}
 Assume $N(\tau)=I$ at $\tau=\pm T/2$ and $\tau=0$. Suppose that the overlaps
  $\langle \psi_k(-\tau)\mid\ \psi_l(\tau)\rangle$ are given by the matrix
 $M(\tau)$:
 \begin{equation}
  M_{k l}(\tau)=\langle \psi_k(-\tau)\mid\ \psi_l(\tau)\rangle ,
 \end{equation}
 so that $M(-\tau)=M(\tau)^+$. The condition $\langle \phi_k(-\tau)\mid
 \phi_l(\tau)\rangle=\delta_{k l}$ means that
 \begin{equation}
   N^+(-\tau)M(\tau)N(\tau) = I  ,
 \end{equation}
 which leads to $M^{-1}(\tau)=N(\tau) N^+(-\tau)$. Calculate action
 in terms of states $\psi_k(\tau)$. The integrand is:
 \begin{eqnarray}
  \label{traces}
  \sum_{i k l} N^*_{ki}(-\tau)\langle \psi_k(-\tau)\mid \partial_{\tau}
 [N_{li}(\tau) \psi_l(\tau)]\rangle  & =   \\ \nonumber
   \sum_{k l} M^{-1}_{l k}(\tau)\langle\psi_k(-\tau)\mid\partial_{\tau}\psi_l(\tau)\rangle
 +  \sum_{i l} N^{-1}_{i l}(\tau)(\partial_{\tau} N_{li}(\tau))   .\\ \nonumber
\end{eqnarray}
 The second term is just:
 $TrN^{-1}\partial_{\tau}N=\partial_{\tau}(\ln \det N)$.
 From Eq.~(\ref{SR}) one obtains:
 \begin{equation}
  \label{Sinv2}
  S/\hbar= 2\Re\int_{0}^{T/2}d\tau
  \sum_{k l} M^{-1}_{l k}(\tau)\langle\psi_k(-\tau)\mid\partial_{\tau}\psi_l(\tau)\rangle ,
 \end{equation}
 where the omitted residual term, $\Re\int_{-T/2}^{T/2}d\tau\partial_{\tau}
 (\ln\det N(\tau))$ is identically zero, and the integration
 interval may be reduced to $[0,T/2]$ due to the properties of $M(\tau)$.
 Expanding either
  $\partial_{\tau}\psi_k(\tau)$ or $\psi_k(-\tau)$ onto $\psi_l(\tau)$ and
 a component perpendicular to all $\left\{ \psi_k(\tau)\right\}_{k=1}^N$,
 one can notice that only the part $[\partial_{\tau}\psi_k(\tau)]_{\perp}$
   orthogonal to all $\psi_l(\tau)$ contributes to action:
  $S/\hbar= 2\Re\int_{0}^{T/2}d\tau
  \sum_{k l} M^{-1}_{l k}(\tau)\langle[\psi_k(-\tau)]_{\perp}\mid
  [\partial_{\tau}\psi_l(\tau)]_{\perp}\rangle$.

 The Slater determinants $\mid \Psi(\tau)\rangle$, built out of
 $\psi_k(\tau)$, are related to bounce determinant states
 $\mid \Phi(\tau)\rangle$ by:
  $\mid\Phi(\tau)\rangle=\det N(\tau) \mid\Psi(\tau)\rangle$,
  so that $\langle\Psi(-\tau)\mid \Psi(\tau)\rangle=\det M(\tau)$ and
 ${\cal H} = \langle\Phi(-\tau)\mid {\hat H} \mid\Phi(\tau)\rangle=
 \langle\Psi(-\tau)\mid {\hat H} \mid\Psi(\tau)\rangle/
 \langle\Psi(-\tau)\mid \Psi(\tau)\rangle$.
  Therefore, energy overlap kernel ${\cal H}$, like action,
 does not involve $N(\tau)$ alone and may be expressed as \cite{RY}:
 \begin{equation}
  \label{hfe}
  {\cal H}=\sum_{i} \langle \psi_i(-\tau)\mid{\hat t}\mid\psi'_i(\tau)\rangle
  +\frac{1}{2}\sum_{i,j} \langle \psi_i(-\tau)\psi_j(-\tau)\mid{\hat v}\mid
 \psi'_i(\tau)\psi'_j(\tau)-\psi'_j(\tau)\psi'_i(\tau)\rangle ,
  \end{equation}
 where the states $\psi'(\tau)$ are related to $\psi(\tau)$ via:
$\psi'_i(\tau)=\sum_k M^{-1}_{k i}(\tau) \psi_k(\tau)$. The s.p. Hamiltonian
 ${\hat h}$ may be expressed in terms of various densities
   which do not involve $N(\tau)$ either, as for example
 $\rho(\tau)=\sum_i\psi^*_i(-\tau)\psi'_i(\tau) =
 \sum_k\mid\psi_k(\tau)\mid^2+\sum_{k l}M^{-1}_{k l}(\tau)
 [\psi^*_l(-\tau)]_{\perp}\psi_k(\tau)$, etc.
  The Eqs.~(\ref{eqsp}) do involve $N(\tau)$:
\begin{equation}
 \label{eqsp2}
\hbar\partial_{\tau}\psi_k+ {\hat h}\psi_k +
 \sum_l \hbar\left[(\partial_{\tau}N) N^{-1}\right]_{l k}\psi_l
  -\sum_l\left[\sum_m
  N_{l m} \epsilon_m N^{-1}_{m k} \right]\psi_l = 0 ,
\end{equation}
 but become independent of it when projected onto a space orthogonal to all
 $\left\{\psi_k(\tau)\right\}_{k=1}^N$:
\begin{equation}
 \label{eqperp}
 \left( \hbar\partial_{\tau}\psi_k(\tau) +
  {\hat h}(\tau)\psi_k(\tau)\right)_{\perp} = 0 ,
\end{equation}
  and only this part is relevant for action.

 When the transformation $N(\tau)$ has the property of a "generalized unitarity",
  $N^{-1}(\tau)=N^+(-\tau)$,
 the overlaps of states $\psi_k(-\tau)$ and $\psi_l(\tau)$ have canonical
 form $M^{-1}(\tau)=N(\tau)N^+(-\tau)=I$. Then each of the matrices $N\epsilon N^{-1}$
 and $(\partial_{\tau}N)N^{-1}$ has a hermitean component which is $\tau$-even and
 an antihermitean component which is $\tau$-odd. For an arbitrary nonsingular $N(\tau)$,
 in particular, such that keeps states $\psi_k(\tau)$ orthonormal, the matrix
 $M(\tau)$ in general depends on $\tau$ and has no defined $\tau$-parity.
 Conversly, a general form of the instanton equation:
 \begin{equation}
 \hbar\partial_{\tau}\psi_k(\tau)+ {\hat h}(\tau)\psi_k(\tau)+
 \sum_l{\cal E}_{l k}(\tau)\psi_l(\tau) = 0 ,
 \end{equation}
preserves overlaps Eq.(\ref{overdel}) if ${\cal E}(\tau)$ has a hermitean
 $\tau$-even and antihermitean $\tau$-odd parts. There is a great variety of
possible instanton representations with different overlaps
 $\langle\psi_k(\tau)\mid\psi_l(\tau)\rangle$ corresponding to different
 matrices ${\cal E}$. The periodicity condition for instanton imposes integral
 conditions: $\int_{-T/2}^{T/2}d\tau\partial_{\tau}(\langle\psi_k(\tau)\mid
 \psi_l(\tau)\rangle)=0$, i.e. integral relations between the matrix elements
 of ${\hat h}_R$, ${\cal E}$ and the overlaps
 $\langle\psi_k(\tau)\mid\psi_l(\tau)\rangle$:
 \begin{equation}
 \label{integra}
  \int_{-T/2}^{T/2}\left( 2\langle\psi_k(\tau)\mid{\hat h}_R(\tau)\mid
 \psi_l(\tau)\rangle+\sum_m ({\cal E}^*_{m k}(\tau)\langle\psi_m(\tau)\mid
 \psi_l(\tau)\rangle+\langle\psi_k(\tau)\mid\psi_m(\tau)\rangle
  {\cal E}_{m l}(\tau))\right) = 0 .
 \end{equation}
 To assure orthonormal $\{\psi_k\}$ at $\tau=0$, both sets of integrals, 
 $\int_0^{T/2}$ and $\int_{-T/2}^0$, should be zero.
 From Eqs.(\ref{eqsp}) we know that ${\cal E}_{k l}=-\epsilon_k\delta_{k l}$
 provides one of the possible choices, but obviously there are many others,
 among them those with the diagonal matrix ${\cal E}$, i.e. with some
  $\tau$-dependent s.p. energies $\epsilon_k(\tau)$.

 For representations with orthonormal s.p. states $\psi_k(\tau)$,
 like for the usual HF determinants, Eqs.(\ref{overdel}) do not hold,
 while the following relations are satisfied:
 $\psi_k(-\tau)=\sum_i M^*_{k i}(\tau) \psi_i(\tau) +
  [\psi_k(-\tau)]_{\perp}$,  and
   $\langle[\psi_k(-\tau)]_{\perp}\mid[\psi_l(-\tau)]_{\perp}\rangle =
  \delta_{k l} - (M(\tau)M^+(\tau))_{k l}$.
 Among them exists a special representation for which
 $\langle\psi_k(\tau)\mid\partial_{\tau}\psi_l(\tau)\rangle = 0$, which means
 that $\partial_{\tau}$ as an operator has only particle-hole (p-h) matrix
 elements. This corresponds to the matrix ${\cal E}$ which fulfils the
 equality ${\cal E} = -{\hat h}$ on the subspace spanned by
  $\left\{ \psi_k(\tau)\right\}_{k=1}^N$.

\section{Variational principle for bounce action}

 Consider variation of action $S$ in terms of some trial fission path defined
 in terms of s.p. states $\phi_{1 k}(\tau)$ and $\phi_{2 k}(\tau)$
 for $0 < \tau < T/2$ as in Eq.(\ref{path}), fulfilling instanton boundary
 conditions
 \begin{eqnarray}
 \delta(S/\hbar) & = & \sum_k\int_{-T/2}^{T/2}\left(\langle
 \delta\phi_k(-\tau)\mid\partial_{\tau}\phi_k(\tau)\rangle-
 \langle\partial_{\tau}[\phi_k(-\tau)]\mid\delta\phi_k(\tau)
 \rangle\right)
 \\ \nonumber
   & = & \sum_k\int_{0}^{T/2}\left(\langle\delta\phi_{1 k}(\tau)\mid
 \partial_{\tau}\phi_{2 k}(\tau)\rangle-\langle\partial_{\tau}
 \phi_{1 k}(\tau)\mid\delta\phi_{2 k}(\tau)\rangle\right) + c.c.
 \end{eqnarray}
 If the states $\phi_{2 k}$ fulfil the first set of Eqs.(\ref{eqpa}) with
 $\phi_{1 k}$ taken as the bra, then
 \begin{equation}
 \delta S  = \sum_k\int_{0}^{T/2}\left(\langle\delta\phi_{1 k}(\tau)\mid
 \epsilon_k-{\hat h}(\tau)\mid\phi_{2 k}(\tau)\rangle-\langle\hbar
 \partial_{\tau}
 \phi_{1 k}(\tau)\mid\delta\phi_{2 k}(\tau)\rangle\right) + c.c.
 \end{equation}
  If, additionally, energy is kept constant so that variations fulfil:
 \begin{equation}
 \delta{\cal H} = \sum_k\left(\langle\delta\phi_{1 k}
(\tau)\mid{\hat h}(\tau)\phi_{2 k}(\tau)\rangle+\langle{\hat h}(-\tau)
 \phi_{1 k}(\tau)\mid\delta\phi_{2 k}(\tau)\rangle\right) = 0,
 \end{equation}
 then, since $\langle\phi_{1 k}(\tau)\mid\phi_{2 k}(\tau)\rangle=1$, variation
 of $S$ reads:
 \begin{equation}
 \label{varS}
 \delta S  = \sum_k\int_{0}^{T/2}\langle({\hat h}(-\tau)-\epsilon_k)
 \phi_{1 k}(\tau)-\hbar\partial_{\tau}\phi_{1 k}(\tau)
 \mid\delta\phi_{2 k}(\tau)\rangle + c.c.
 \end{equation}
 As may be seen from this equation, after the first set of Eqs.(\ref{eqpa}) and
  the condition ${\cal H}=E_{gs}$ are fulfilled, action $S$ ceases to be a
 functional of both $\phi_{1 k}$ and $\phi_{2 k}$ and becomes a functional
 of $\phi_{2 k}$ and their time derivatives $\partial_{\tau}\phi_{2 k}$.
 The functions $\phi_{1 k}$ provide, through the s.p. Hamiltonian, the drive
 for $\phi_{2 k}$ which is exactly required to produce
 $\partial_{\tau}\phi_{2 k}$.
 As we have argued in section II, and as follows from the physical meaning of
 action, for such a driven motion $S$ must be positive.
 Since $\delta S[\phi_{2 k}]$ Eq.(\ref{varS}) vanishes for
 $\phi_{1 k}$ that fulfil the second set of instanton equations (\ref{eqpa}),
  i.e. when $\phi_{1 k}$ and $\phi_{2 k}$ together form instanton,
 the instanton action must be a minimum of $S[\phi_{2 k}]$.
 Thus, for $\phi_{2 k}$ and $\phi_{1 k}$
 such that: both fulfil the instanton boundary conditions, the overlap
 condition Eq.(\ref{overdel}), the energy condition
 ${\cal H}=E_{gs}$ and $\phi_{1 k}$
 solve the first set of Eqs.(\ref{eqpa}) for $\partial_{\tau}\phi_{2 k}$ -
 calculated action provides an upper bound for action of the optimal
 (i.e. the one with the smallest action, if there are a few)
  instanton.
 That the last condition is necessary may be seen from the negative sign of
 action for bounces evolving backwards in $\tau$, Eq.(\ref{SR}).
 The assumption that half of the bounce equations are fulfilled
 eliminates trial paths with admixtures of instantons evolving backwards
 which would leave the sign of action undecided. (Note, that action
 for instanton evolving backwards attains the maximal among negative
 values.)
 This is in complete analogy to mechanics, where
  the real motion $(q_i,{\dot q}_i)$ minimizes action
  $\int \sum_i p_i dq_i$  under the condition of constant
  energy {\it provided} that canonical
 relations ${\dot q}_i=\partial{\cal H}/\partial p_i$ are satisfied on each
 path. The variables introduced in the next section will make this analogy
 even closer.

 One can use the principle of minimal action for any
 representation of a trial path. A simple choice is to take for
 $\psi_{2 k}$ some orthonormal HF states with the proper boundary conditions
 and to look for such $[\psi_{1 k}]_{\perp}$ that
 $\psi_{1 k}=\psi_{2 k}+[\psi_{1 k}]_{\perp}$ fulfil Eqs.(\ref{eqperp}) with
  some $\tau$-reparametrization like in Eqs.(\ref{eqrep})
 \begin{equation}
 \label{eqSmin}
 \left({\dot \theta} \partial_{\theta}\psi_{2 k}+
 {\hat h}[\psi_{1 k},\psi_{2 k}]\psi_{2 k}\right)_{\perp} = 0 .
\end{equation}
 In this representation the overlap conditions are automatically fulfilled.
 Leaving $\tau$-reparametrization free one gains a parameter ${\dot \theta}$
 that allows to control bounce velocity, i.e. the energy condition.
 One can decompose the s.p. mean-field Hamiltonian as suggested by the formula
 for density $\rho$ preceding Eqs.(\ref{eqsp2}),
 ${\hat h}[\psi_{1 k},\psi_{2 k}]={\hat h}[\psi_{2 k}]+\Delta{\hat V}
 [[\psi_{1 k}]_{\perp},\psi_{2 k}]$, with ${\hat V}$ the s.p. potential,
  so that the equation for $[\psi_{1 k}]_{\perp}$ becomes
 \begin{equation}
 \label{eqSmin1}
-\left({\dot \theta} \partial_{\theta}\psi_{2 k}+
 {\hat h}[\psi_{2 k}]\psi_{2 k}\right)_{\perp} =
 \left(\Delta{\hat V}[[\psi_{1 k}]_{\perp},\psi_{2 k}]\psi_{2 k}\right)
 _{\perp} .
\end{equation}
 For complex wave functions, Eqs.(\ref{eqSmin1}) should be solved together
 with their complex conjugate for both $[\psi_{1 k}]_{\perp}$ and
  $[\psi^*_{1 k}]_{\perp}$.
 For small $[\psi_{1 k}]_{\perp}$, one could expand the r.h.s. of
 this equation to linear terms in particle-hole components
 $[\psi_{1 \perp}]_{ph}$ with respect to $\{\psi_{2 h}\}$,
 $[\psi_{1 h}]_{\perp}=\sum_p[\psi_{1 \perp}]_{p h}\mid p>$,
   and try to solve the system of linear equations with the matrix:
 $\partial[\Delta{\hat V}\psi_{2 h}]_{p}/\partial[\psi^*_{1 \perp}]_{p'h'}$.
 This matrix, $\langle p p'\mid {\hat v}\mid {\widetilde {h h'}}\rangle$,
 where tilde means antisymmetrization, is the off-diagonal block of the 
 RPA matrix (with respect to the HF state built of $\{\psi_{2h}\}$),
 which also appears in the ATDHF, cf Eqs. (2.25-2.29) and (8.24) in 
 \cite{BV78}, also \cite{GQ80}.
 The solution of Eqs. (\ref{eqSmin1}) should be obtained for many velocities
  ${\dot \theta}$ to find the one which matches the energy condition.
  For larger barriers, larger
 differences between $\psi_{1 k}$ and $\psi_{2 k}$ are necessary to
 lower the energy overlap kernel ${\cal H}$ to $E_{gs}$. Then, the solution to
  Eqs.(\ref{eqSmin}) or (\ref{eqSmin1}) beyond the
   linear limit does not seem trivial.
 However, if found by any means, it provides action $S$ being an upper bound
 for the decay exponent.

\section{Instantons in coordinate-momentum variables}

 There are natural choices of instanton variables that correspond to
 time-even coordinates and time-odd momenta.
 One possibility is given by \cite{JS1}:
 $\phi_k(\tau)=\varphi_k(\tau)-\xi_k(\tau)$,
 $\phi_k(-\tau)= \varphi_k(\tau)+\xi_k(\tau)$.
 It follows that $\varphi_k(-\tau)=\varphi_k(\tau)$ and $\xi_k(-\tau)=
 -\xi_k(\tau)$.
 Due to the boundary conditions, $\varphi_k(\pm T/2)=\psi_k^{HF}$,
 $\varphi_k(0)= \phi_k(0)$, $\xi_k(\pm T/2)= \xi_k(0)=0$. Thus,
 $\varphi_k$ are average tunneling states (coordinates) which
 may be parametrized by some deformation $Q(\tau)$ (or its real part,
 cf section II), so that $\partial_{\tau}\varphi_k
 = {\dot Q}\partial_Q\varphi_k$.
 The $\tau$-odd components $\xi_k$ must be proportional to $\tau$-odd
 derivative ${\dot Q}(\tau)$, i.e. to collective velocity.
 These two sets of states fulfil the system of equations:
 \begin{equation}
  \label{for2}
 \hbar\frac{\partial}{\partial \tau} \left( \begin{array}{c}
  \varphi_k \\ \xi_k \end{array} \right) = \left(
  \begin{array}{ccc}
  - {\hat h}_A & , & {\hat h}_R -\epsilon_k \\
   {\hat h}_R -\epsilon_k & , & - {\hat h}_A
  \end{array}
  \right) \left( \begin{array}{c} \varphi_k \\ \xi_k \end{array} \right) ,
 \end{equation}
 where we have used decomposition ${\hat h}(\tau)={\hat h}_R+{\hat h}_A$.
 These equations may be obtained either by decomposing Eqs.(3) or
 by deriving equations of motion from the functional
 $\int d\tau\langle\Phi(-\tau)\mid \hbar\partial_{\tau}+{\hat H}\mid\Phi(\tau)
 \rangle$ expressed by $\varphi_k$ and $\xi_k$.
 In the latter case, one has to remember that $\varphi_k(-\tau)$ and
 $\xi_k(-\tau)$ no longer exist as independent variables. The canonical form
 of Eqs.(\ref{for2}), without the periodicity-fixing terms, is
 \begin{eqnarray}
 \label{canon}
 \hbar\frac{\partial\varphi_k(\tau)}{\partial{\tau}}&=& -\frac{\delta{\cal H}}
 {\delta\xi_k^*(\tau)}, \\  \nonumber
 \hbar\frac{\partial\xi_k(\tau)}{\partial{\tau}}&=& \frac{\delta{\cal H}}
 {\delta\varphi_k^*(\tau)}, \\  \nonumber
 \end{eqnarray}
 with canonical pairs $(\varphi_k,\xi_k^*)$ and $(\varphi_k^*,\xi_k)$.
  Densities may be expressed in terms of $\varphi_k$ and $\xi_k$, for example,
  one has: $\rho(x)=\sum_k(|\varphi_k(x)|^2-|\xi_k(x)|^2-
 2i\Im(\varphi_k^*(x)\xi_k(x)))$, etc.
 The conserved overlaps in terms of the amplitudes
 $\varphi_k$ and $\xi_k$ read:
 \begin{eqnarray}
 \label{oversym}
 \langle \varphi_k\mid\varphi_l\rangle-\langle\xi_k\mid\xi_l\rangle &=&
 \delta_{kl}  , \\ \nonumber
 \langle\varphi_k\mid\xi_l\rangle-\langle\xi_k\mid\varphi_l\rangle & = & 0 .
  \\  \nonumber
 \end{eqnarray}

 The first set of Eqs.(\ref{for2}) is consistent with
 $\xi_k$ being proportional to the collective velocity
 ${\dot Q}(\tau)$. In particular, ${\hat h}_A$ contains
  $\xi_k$ in odd orders, for example, the antihermitean
 component of the part $(-{\bf j}\cdot\nabla)$ of the Skyrme-type s.p.
 mean field is proportional to a piece
 $-((\xi^*_i\nabla\varphi_i-\varphi^*_i\nabla\xi_i)/2+c.c.)$ of
 the current density ${\bf j}$.
 The adiabatic limit corresponds to small ${\dot Q}$ and thus small
 $\mid\xi_k\mid$.

 It may be seen that the instanton dependence on ${\dot Q}$ allows
 to satisfy the bounce condition at $\tau=0$:
 as $\xi_k={\dot Q}{\bar \xi}_k$ with ${\bar \xi}_k$ $\tau$-even,
 the time derivative in the second set of Eqs.(\ref{for2}),
 ${\ddot Q}{\bar \xi}_k+{\dot Q}^2\partial_Q{\bar \xi}_k$ reduces to
 ${\ddot Q}{\bar \xi}_k$ at $\tau=0$, where ${\dot Q}=0$ (we assume real $Q$).
 Then ${\hat h}_A(0)=0$, so from the first set of Eqs.(\ref{for2}),
 ${\bar \xi}_k(0)=
 ({\hat h}(0)-\epsilon_k)^{-1}\partial_Q\varphi_k(0)$. Substituting this
 to the second set, we obtain the bounce condition at $\tau=0$
\begin{equation}
 \label{bounce}
 {\ddot Q}\frac{\partial \varphi_k}{\partial Q}(0) =
 ({\hat h}(0)-\epsilon_k)^2 \varphi_k(0) ,
 \end{equation}
 where ${\ddot Q}=\frac{1}{2}d{\dot Q}^2/dQ$ is
 negative at $\tau=0$, and ${\dot Q}^2$ is determined as a function of $Q$
 by the energy condition ${\cal H}[\varphi_k(Q),{\dot Q}{\bar \xi}_k(Q)]=
 E_{gs}$. The exact Eqs.(\ref{bounce}) follow from the combined
 Eqs.(\ref{for2}) and therefore should not be imposed on trial paths in
 a variational search for instantons.

  Due to the symmetry properties of the
 amplitudes, action reads
 \begin{equation}
 \displaystyle S/\hbar = 2 \Re \int_{-T/2}^{T/2} d\tau \sum_k
 \left\langle \xi_k {\bigg|} \frac{\partial \varphi_k}{\partial \tau}
 \right\rangle.
 \end{equation}
 In this expression one immediately recognizes the
 familiar form $\int p_i dq_i$.
 The first set of Eqs.(\ref{for2}) are the velocity-momentum relations which
 should be fulfilled on trial trajectories in a search for bounce as
 a minimum of the action functional. Solving formally for momenta $\xi_k$ and
 substituting into action one obtains
 \begin{equation}
 \label{action}
 S  =  2 \hbar \int_{-T/2}^{T/2} d\tau \sum_k
 \left\langle\hbar\frac{\partial \varphi_k}{\partial {\tau}}+{\hat h}_A(\tau)
 \varphi_k {\bigg|}\frac{1}{{\hat h}_R(\tau)-\epsilon_k}
 {\bigg|}\frac{\partial \varphi_k}{\partial \tau}\right\rangle .
 \end{equation}
 Let us compare this formula to a standard treatment of the spontaneous 
 fission, in which one uses a family of static HF states, each constrained 
 to have a prescribed quadrupole moment $q$, with values of $q$ covering the 
 barrier region. In such a study, one has {\it to assume} some form of the mass 
  parameter $M(q)$ that allows to express collective kinetic energy as 
   $\frac{1}{2} M(q){\dot q}^2$ and action 
  as $\int M(q) {\dot q} dq$, with the implicitly understood energy 
  conservation $V(q)-E_{g.s.}= \frac{1}{2}M(q){\dot q}^2$. In the cranking 
  approximation, 
  $M(q)=2\hbar^2\sum_k\langle\partial\psi_k/\partial q\mid 
  ({\hat h}_{ad}(q)-e_k(q))^{-1}\mid\partial\psi_k/\partial q\rangle$, 
  with the adiabatic mean-field Hamiltonian ${\hat h}_{ad}$ and its 
  eigenenergies $e_k$ depending on $q$.
  After introducing a reparametrization $q(t)$ in terms of some 'time' 
  variable $t$ to have the correspondence with Eq.(\ref{action}),   
  action in the cranking approximation can be written as 
 \begin{equation}
  \label{Scra}
  S_{crank}=2\hbar^2\int_{-T/2}^{T/2}dt\sum_k\left\langle\frac{\partial\psi_k}
 {\partial t}{\bigg|}\frac{1}{{\hat h}_{ad}(t)-e_k(t)}{\bigg|}\frac
 {\partial\psi_k}{\partial t} \right\rangle . 
  \end{equation}
  One can see that Eq.(\ref{action}), after neglecting the Thouless-Valatin 
  term, is deceptively similar to the cranking expression.
 (The Thouless-Valatin term changes cranking masses by less than 20\%
 \cite{GQ80II}.) However, a closer look reveals important
  differences: The constants $\epsilon_k$ in the denominator in (\ref{action}) 
  are the s.p. energies at the metastable HF minimum, not the adiabatic  
  eigenenergies $e_k(q(t))$; the states $\varphi_k$, generally not 
  orthonormal, are not equal to the adiabatic s.p. eigenstates $\psi_k(q(t))$; 
 the self-consistent s.p. Hamiltonian in the instanton method 
  depends on $\tau$-odd amplitudes, ${\hat h}={\hat h}[\varphi_k,\xi_k]$, 
 and this requires an iterative solution of the velocity-momentum relations.

 As follows from section III, $\epsilon_k$ could be replaced in the instanton  
 Eqs.(\ref{eqpa},\ref{for2}) by some $\tau$-dependent quantities 
 ${\tilde \epsilon}_k(\tau)$. Such a change results from scaling the s.p.  
  bounce states via  
 $\phi_k(\tau)=\phi_k'(\tau)\exp(\int_0^{\tau}(\epsilon_k-
{\tilde \epsilon}_k(\tau')) d\tau'/\hbar)$, 
 with $\tau$-even  ${\tilde \epsilon}_k$. 
 This is a particular linear transformation of the type (\ref{linear}) 
 which preserves canonical overlaps Eq.(\ref{overdel}) 
 and the periodicity, if the conditions 
$\int_0^{T/2}d\tau\Delta\epsilon_k(\tau)/\hbar=0$ are satisfied  
 with $\Delta\epsilon_k=\epsilon_k-{\tilde \epsilon}_k(\tau)$. 
 After such transformation, $\xi_k=\cosh(y)\xi_k'-\sinh(y)\varphi_k'$
 with $y(\tau)=\int_0^{\tau}d\tau'\Delta\epsilon_k/\hbar$, so both 
 $\xi_k'$ and $y$ have to be of the order 
 ${\dot Q}$ to keep $\xi_k\sim{\dot Q}$ for small ${\dot Q}$. This requires that the 
 average $\Delta\epsilon_k$ be of the order ${\dot Q}^2$, so only a mild 
 deformation-dependence of adiabatic energies is compatible 
  with bounce properties.

 A trial fission path is adiabatic if $\{\varphi_k\}$ differ only a little from 
  orthonormal eigenstates of ${\hat h}_R$ with energies 
 $\tilde \epsilon_k(\tau)$ obtained by such a rescaling, and the 
 velocity-momentum relations produce small $\xi_k$. Then  
   ${\hat h}_R[\varphi_k]$ may be considered the adiabatic mean field and 
  the cranking amplitudes $\xi_k$ solve the second set of Eqs.(\ref{oversym}). 
 This suggests (and will be shown  
  by a different method in the next section) that in the adibatic limit 
  $S_{crank}$ provides an upper bound of (\ref{action}) with the neglected 
  Thouless-Valatin term.   

   Otherwise, when the larger $\xi_k$ are required,    
   the self-consistency and conditions (\ref{oversym}) induce a large 
  difference between the contents of the cranking and instanton-motivated 
  forms of action.  
 For $\xi_k$ not small, the enforcement of the velocity-momentum
  conditions together with Eqs.(\ref{oversym}) seems difficult.
 The same difficulty remains in the action minimization within this 
 representation:
 since the properties of solutions to (\ref{for2}) are not assured for
 trial paths, the conditions for overlaps (\ref{oversym}) should be
 imposed on them independently of other necessary conditions.

 \section{Adiabatic limit of the instanton method}

 A framework analogous to that of the ATDHF theory may be obtained by defining
 other variables. One can observe that, due to the
  overlap conditions (\ref{overdel}), a linear transformation that
  maps each $\phi_k(\tau)$ into $\phi_k(-\tau)$ may be completed
  to a hermitean operator. Denoting the square root of this operator at
  each $\tau$ as $\exp({\hat S}(\tau))$, with ${\hat S}(\tau)$ hermitean,
  we have $\exp(2{\hat S}(\tau))\phi_k(\tau)=\phi_k(-\tau)$ for all
  $\tau$ and $k$. Substituting $-\tau$ for $\tau$ in this relation and
  comparing both, we infer
  that ${\hat S}(-\tau)=-{\hat S}(\tau)$. Then,
 $\exp({\hat S}(\tau))\phi_k(\tau)=\exp({\hat S}(-\tau))\phi_k(-\tau)$ for all
  $\tau$ and $k$. This means that the above defined vectors, which we will
 call $\psi_{0 k}(\tau)$, are time-even and orthonormal.
  Thus we have
 \begin{eqnarray}
 \phi_k(\tau) & = & \exp(-{\hat S}(\tau))\psi_{0 k}(\tau) , \\ \nonumber
 \phi_k(-\tau) &= & \exp({\hat S}(\tau))\psi_{0 k}(\tau) , \\  \nonumber
 \end{eqnarray}
 with $\psi_{0 k}(\tau)$ some $\tau$-even orthonormal states and
 ${\hat S}(\tau)$ a $\tau$-odd operator. The relation of these new
  variables to those from the previous section is given by:
 $\varphi_k=\cosh({\hat S})\psi_{0 k}$ and $\xi_k=\sinh({\hat S})\psi_{0 k}$.
 The condition ${\hat S}^+ ={\hat S}$ ensures the constant overlaps
 Eq.(\ref{overdel}). The bounce boundary conditions in terms of the new
 coordinates read: $\psi_{0 k}(\pm T/2)=
 \psi_k^{HF}$, $\psi_{0 k}(0)=\phi_k(0)$ and ${\hat S}(\pm T/2)={\hat S}(0)=0$.
 The states $\psi_{0 k}$ define a $\tau$-even density matrix analogous to the
 $\rho_0$ of the ATDHF theory \cite{GQ80}. However, the object $e^{-{\hat S}}
 \rho_0 e^{\hat S}$ does not define any density matrix, contrary to
 $e^{i{\hat \chi}}\rho_0 e^{-i{\hat \chi}}$ of the ATDHF. The $\tau$-odd
 matrix ${\hat S}$ must be proportional to ${\dot Q}(\tau)$. It
 introduces time-odd components to the s.p. wave functions and its smallness is
 equivalent to the adiabaticity condition. The instanton equations may be
 written as:
 \begin{equation}
 \label{eqsymS}
 \hbar(e^{\hat S}(\partial_{\tau}e^{-{\hat S}})\psi_{0 k} +
 \partial_{\tau}\psi_{0 k}) + e^{\hat S}({\hat h}(\tau)-\epsilon_k)
 e^{-{\hat S}}\psi_{0 k} = 0 .
 \end{equation}
 Using expansions (with any operator ${\cal O}$):
\begin{eqnarray}
 e^{\hat S}{\cal O}e^{-{\hat S}}&=&{\cal O} + [{\hat S},{\cal O}] + \frac{1}{2!}
 [{\hat S},[{\hat S},{\cal O}]] +
 \frac{1}{3!}[{\hat S},[{\hat S},[{\hat S},{\cal O}]]] + ... \\   \nonumber
 e^{\hat S}(\partial_{\tau}e^{-{\hat S}})&=&-\left(\partial_{\tau}{\hat S}
  + \frac{1}{2!}[{\hat S},\partial_{\tau}{\hat S}] +
 \frac{1}{3!}[{\hat S},[{\hat S},\partial_{\tau}{\hat S}]] + ...\right)
  \\ \nonumber
\end{eqnarray}
one can split Eq.(\ref{eqsymS}) into $\tau$-even and $\tau$-odd parts.
 So obtained equations are exact when the full expansion is kept. Since
 ${\hat h}={\hat h}[e^{\hat S}\psi_{k0},e^{-{\hat S}}\psi_{k0}]$,
  ${\hat h}_R(\tau)$ contains all even, and ${\hat h}_A(\tau)$ all odd
  orders of ${\hat S}$. The approximation valid to the $n$-th order in
 ${\hat S}$ consists in keeping the appropriate number of terms in both 
 ${\hat h}_R$ and ${\hat h}_A$ in each term of the equations. 

 In the adiabatic limit one expects that the time derivative introduces one 
 order of smallness, so, for example, $\partial_{\tau}{\hat S}$ is of the 
 order of ${\hat S}^2$.  Then, up to the terms of the second order in 
 ${\hat S}$ the equations read
 \begin{eqnarray}
 \label{eqS}
 \left({\hat h}_R - \epsilon_k - \hbar\partial_{\tau}{\hat S} +
 \frac{1}{2}[{\hat S},[{\hat S},{\hat h}_0]] + [{\hat S},{\hat h}_A]\right)
 \psi_{0 k}& =& 0  ,\\  \nonumber
 \hbar\partial_{\tau}\psi_{0 k} + \left([{\hat S},{\hat h}_0] 
 + {\hat h}_A \right)\psi_{0 k}& =& 0 ,
 \\ \nonumber
 \end{eqnarray}
 with the first order ${\hat h}_A$, and ${\hat h}_R$ of the order zero, 
 equal to ${\hat h}_0={\hat h}[\psi_{0 k}]$, except for the first term of 
 the first equation, where the second order 
 ${\hat h}_R$ should be used. In the time-odd equation, the lacking terms 
 start at the order three, and would include 
 $-\frac{\hbar}{2}[{\hat S},\partial_{\tau}{\hat S}]\psi_{0 k}$, etc. 
 As discussed in the previous section, the difference between 
 constants $\epsilon_k$ and the adiabatic energies $\epsilon_k(\tau)$, 
 which may be understood as the expectation values  
  $\langle\psi_{0 k}\mid{\hat h}_0\mid\psi_{0 k}\rangle$, resides in the 
 diagonal part of $\partial_{\tau}{\hat S}$, 
 generically of the order ${\dot Q}^2$. Clearly, not every static HF path 
 is a proper candidate for $\tau$-even bounce components $\psi_{0 k}$, 
 even if bounce is adiabatic (i.e. ${\hat S}$ is small). 
   
 In terms of $\psi_{0 k}$ and ${\hat S}$ action is given by 
 \begin{equation}
 S/\hbar = \Re\int_{-T/2}^{T/2}\sum_k 
 \langle\psi_{0 k}\mid e^{\hat S}(\partial_{\tau} e^{-{\hat S}})
 \mid\psi_{0 k}\rangle , 
  \end{equation}
 as the part of the integrand involving 
 $\partial_{\tau}\psi_{0 k}$ is identically zero due to the normalization of 
  $\psi_{0 k}$.  

 The approximation analogous to the ATDHF consists in solving the second
 Eq.(\ref{eqS}) up to the first order in ${\hat S}$.
 With a given Hamiltonian, energy up to the second order in ${\hat S}$
 reads ${\cal H}_0 +
 \frac{1}{2}\langle\Psi_0\mid[{\hat S},[{\hat S},{\hat H}]]\mid\Psi_0\rangle$,
 with ${\cal H}_0 = \langle\Psi_0\mid{\hat H}\mid\Psi_0\rangle$.
 The term quadratic in ${\hat S}$ is negative and equal to $Tr(\rho_0
 [{\hat S},[{\hat S},{\hat h}_0]+{\hat h}_A])/2$, with ${\hat h}_0={\hat h}
 [\rho_0]$ and ${\hat h}_A$ linear in ${\hat S}$. The latter operator 
  is defined through its matrix elements between arbitrary states 
  $\mid\alpha\rangle$ and $\mid\beta\rangle$:
 \begin{equation}
  \langle \alpha\mid{\hat h}_A\mid\beta\rangle=
  \sum_k\left( \langle \alpha ({\hat S}\psi_{0 k})\mid{\hat v}\mid
  {\widetilde {\beta \psi_{0 k}}}\rangle-
  \langle \alpha \psi_{0 k}\mid{\hat v}\mid
  {\widetilde {\beta({\hat S}\psi_{0 k})}}
  \rangle \right) ,
 \end{equation}
 with tilde denoting antisymmetrization.
 Up to the second order in ${\hat S}$, action is given by 
  $S/\hbar = -\Re\int_{-T/2}^{T/2}\sum_k 
 \langle\psi_{0 k}\mid \partial_{\tau} {\hat S}\mid\psi_{0 k}\rangle$, 
 which may be expressed as 
 \begin{equation}
 S/\hbar = 2\Re\int_{-T/2}^{T/2}\sum_k \langle\psi_{0 k}\mid{\hat S}
 \mid\partial_{\tau}\psi_{0 k}\rangle  .
 \end{equation}
 The lacking terms start at the order four, as the contribution of the 
 order three, with the time-odd integrand 
 $-Tr(\rho_0[{\hat S},\partial_{\tau}{\hat S}])/2$, vanishes.  
 After using the second Eq.(\ref{eqS}), action in the adiabatic 
 limit reads    
 \begin{equation}
 \label{actadia}
  S=-\int_{-T/2}^{T/2} d\tau \sum_k\left(\langle\psi_{0 k}\mid {\hat S}
 ({\hat h}_A+[{\hat S},{\hat h}_0])\mid\psi_{0 k}\rangle + c.c.
  \right) \\ =-\int_{-T/2}^{T/2} d\tau \sum_k
 \langle\psi_{0 k}\mid [{\hat S},{\hat h}_A+[{\hat S},{\hat h}_0]]\mid
 \psi_{0 k}\rangle ,
 \end{equation}
 and hence is equal to the integral of:
 $-\langle\Psi_0\mid[{\hat S},[{\hat S},{\hat H}]]\mid\Psi_0\rangle=
 -2({\cal H}-{\cal H}_0)$.

 If one has an energy functional instead of Hamiltonian, one still 
 obtains action (\ref{actadia}). 
 The integrand may be shown equal to 
  $-2({\cal H}[\varphi_k,\xi_k]-{\cal H}[\psi_{0 k}])$, with $\varphi_k=
 (1+{\hat S}^2/2)\psi_{0 k}$ and $\xi_k={\hat S}\psi_{0 k}$: 
 One calculates 
  $\delta {\cal H}={\cal H}[\psi_{0 k}+\delta\varphi_k,\xi_k+\delta\xi_k]-
   {\cal H}[\psi_{0 k},\xi_k]$ for $\delta \varphi_k={\hat S}^2\psi_{0 k}/2$,
  $\xi_k={\hat S}\psi_{0 k}$ and $\delta \xi_k$ smaller than $\xi_k$, 
  to the second order in ${\hat S}$ by using Eqs. (\ref{for2}),(\ref{canon}):  
 \begin{equation}
  \delta{\cal H} = \sum_k\left(\langle\delta\xi_k\mid {\hat h}_A(\tau)\mid
 \varphi_k\rangle -\langle\delta\xi_k\mid {\hat h}_0(\tau)\mid\xi_k\rangle
  +\langle\delta\varphi_k\mid {\hat h}_0(\tau)\mid\varphi_k\rangle +c.c.
  \right) .
 \end{equation}
 Then one deduces
  $\delta(\sum_k \langle\xi_k\mid {\hat h}_A\mid\varphi_k\rangle+c.c.)= 2
 (\sum_k\langle\delta\xi_k\mid {\hat h}_A\mid\varphi_k\rangle+c.c.)$ 
  and $\delta\langle\xi_k\mid{\hat h}_R\mid\xi_k\rangle=
 (\langle\delta\xi_k\mid{\hat h}_0\mid\xi_k\rangle+c.c)$ at  
 the second order in ${\hat S}$.
 Thus, either with the Hamiltonian or the density functional, one obtains 
 the same form of the positive integrand, which, when presented as
 ${\dot Q}^2\times mass$, defines a positive mass for tunneling.

  In ATDHF, the linear response limit of the time-odd equation, i.e. the 
  counterpart of the second Eq.(\ref{eqS}), is: 
  $\hbar\partial_{\tau}\psi_{0 k}+ (i{\hat h}_1+
  [{\hat \chi},{\hat h}_0])\psi_{0 k}=0$, with 
  ${\hat h}_0={\hat h}[\psi_{0 k}]$,
 ${\hat h}_1= iTr_2({\tilde v}[{\hat \chi},\rho_0])$, ${\tilde v}$ 
 the antisymmetrized interaction and $Tr_2$ meaning trace over coordinates of 
 the second particle. 
  However, ${\hat h}_1=-i{\hat h}_A({\hat \chi})$, so that the $\tau$-odd 
  equation for the instanton operator ${\hat S}$ is a copy of the ATDHF 
  equation, with ${\hat S}={\hat \chi}$.
  Thus, in the adiabatic limit, instanton action defines the ATDHF mass
  $\hbar Tr({\hat S}{\dot \rho}_0)/{\dot Q}^2$.
  In both cases only the particle-hole components of ${\hat S}$ are
  determined. 

  The first Eq.(\ref{eqS}) provides the adiabaticity
  condition for a trial path, as in ATDHF \cite{BV78,GQ80}, but with a
  different sign by $\partial_{\tau}{\hat S}$. It is worth
  emphasizing though, that this condition was practically never checked
   in calculations of ATDHF masses. Thus, up to now, decay probabilities
  were calculated without knowing whether a chosen fission path is compatible
  with this equation.
   As far as action is concerned, the difference between the real- and
  imaginary-time dynamics, i.e. between
  oscillations and tunneling, appears in the next order.

  A search for instanton in the adiabatic limit would consist in looking
  for the minimum of action determined by the ATDHF mass over trial paths
  that should fulfil the adiabaticity condition.
  It is well known that near the s.p. level crossing at the Fermi
  surface, an extremely small velocity is needed to keep the
  occupation of the lower level. Since in ATDHF ${\dot Q}$
  must be also adjusted to keep the bulk energy ${\cal H}$ constant, it may
  fail to fulfil two requirements simultaneously in the vicinity
  of the crossing.
  Thus, the proper ATDHF fission path should avoid such crossings.
  Fission paths that break many symmetries, along which crossings are avoided
  by a strong interaction between levels, could provide one remedy
  for this problem (as suggested by the calculations reported in \cite{JN}).
  The other would be to solve
  Eq.(\ref{eqS}) for instanton to the higher order in ${\hat S}$, which would
  modify the mean field ${\hat h}_R$ and avoid crossings present for
  the initial ${\hat h}_0$. Finally, a partial remedy is given by 
  pairing.

 \section{Bounce action vs GCM inertia}

 The use of the variational principle for instantons depends on the
 ability to impose the velocity-momentum conditions.
 These conditions are crucial, as without them
  action for a trial path may be lower than that for bounce.
  Below, we show that the GCM formula for a collective mass that restricts
  generating states to $\tau$-even Slater determinants respects only
 the energy condition and hence is incompatible with the instanton
  method.

  Consider a family of orthonormal states labeled by the quadrupole moments
 $q_1(\tau)$ and $q_2(\tau)$, $\tau>0$, and calculate action Eq.(\ref{Sinv2}).
 Through the barrier, $q_2(\tau)$ must be different from $q_1(\tau)$ to make
  energy overlap kernel
 $\langle \Psi(q_1(\tau))\mid {\hat H} \mid \Psi(q_2(\tau))\rangle/
  \langle\Psi(q_1(\tau))\mid \Psi(q_2(\tau))\rangle$ equal to $E_{gs}$.
 If we suppose that $\Psi$ depends solely on $q$ and not on ${\dot q}$, as in
 many GCM studies, the matrix $M(\tau)$ becomes a function of $q_1$ and $q_2$,
 the integrand in Eq.(\ref{Sinv2}) becomes equal to
  $Tr (M(q_1,q_2)^{-1}(\partial M(q_1,q_2))/\partial{q_2})$ and
    \begin{equation}
     \label{Sq}
     S = 2 \hbar\Re \int_{q(0)}^{q(T/2)} dq_2
    \frac{\partial \ln\det M(q_1,q_2)}{\partial q_2} .
    \end{equation}
   From this equation one can deduce a connection between the signs of
   $S$ and $q_2-q_1$:
  The Eqs.(\ref{eqsp}) and (\ref{eqpa}) tell that the deformation $q_1$ of the
  state $\Psi(q_1)$ drags deformation $q_2$ of $\Psi(q_2)$, thus $q_2$ lags
  behind $q_1$ on the way from behind the barrier to the metastable
  minimum, i.e. $q_2(\tau)>q_1(\tau)$. Therefore, increasing $q_2$
 while keeping $q_1$ fixed {\it increases} separation between $q_1$ and $q_2$,
  and thus decreases the overlap $\det M(q_1,q_2)$.
  Hence, the integrand in (\ref{Sq}) is negative, as is differential
  $dq_2$ (as $q(0)>q(T/2)$), so action $S$ is positive.

  In the above reasoning we used the property of the bounce equation. While
  using variational principle, one might exchange the states $\Psi_1$ and
  $\Psi_2$, and then, by the previous reasoning, negative action would follow.
  One might try to take $\mid S\mid$ for action in such a case, and there
  are cases in which this way of proceeding defines a minimum.
  At the same time, it is clear that some additional conditions are
  necessary in the variational formulation.

 One can expand the integrand in Eq.(\ref{Sq}) with respect
 to the quadrupole moment difference $s=q_2(\tau)-q_1(\tau)$ around the
 midpoint ${\bar q}=(q_1+q_2)/2$. When one assumes the
 GOA: $\ln\det M(q_1,q_2)\approx -\gamma({\bar q})s^2/2$, and
 then disregards quadratic and higher order terms in $s$,
 one obtains:
 \begin{equation}
    S\approx -2\hbar \int_{q(0)}^{q(T/2)} dq_2 \gamma({\bar q})(q_2-q_1) ,
  \end{equation}
 where, as discussed above, $q_2>q_1(q_2)$, and $\gamma({\bar q})=
 \sum_k\langle\partial_q\psi_k\mid\partial_q\psi_k\rangle-
 \sum_{k l}\langle\partial_q\psi_k\mid\psi_l\rangle\langle\psi_l\mid\partial_q
 \psi_k\rangle$. The integration variable $dq_2=d{\bar q}+ds/2$ may be changed
 to $d{\bar q}$, as the integral $sds=d(s^2)/2$ between the endpoints with
 $s=0$ vanishes.
 The difference of the quadrupole moments may be calculated from the
 constraint on the energy overlap kernel:
$E_{gs}={\cal H}[q_2,q_1]\approx {\cal H}[{\bar q},{\bar q}]-s^2({\cal H}_{xy}
-{\cal H}_{xx})/4$,
 where we have used the symbolic notation for derivatives of ${\cal H}$,
 e.g. ${\cal H}_{xx}=\partial_x^2{\cal H}(x,y)\mid_{x=y={\bar q}}$, etc.,
 and conditions ${\cal H}_x={\cal H}_y$, ${\cal H}_{xx}={\cal H}_{yy}$ holding 
  for time-even ${\cal H}$ 
 (cf. \cite{RingS}, where the discussion of those is given).
 Since the diagonal value of the energy overlap is just
 "potential energy" $V({\bar q})$ in the standard approach, we obtain:
 \begin{equation}
    S\approx 2\hbar \int_{q(T/2)}^{q(0)} d{\bar q} \sqrt
  {2\left(V({\bar q})-E_{gs}\right)\left(\frac{2\gamma({\bar q})^2}{{\cal H}_{xy}-{\cal H}_{xx}}\right)} ,
 \end{equation}
 where the quantity in the second parenthesis under the square root sign
 is the GCM+GOA mass (cf. \cite{RingS}).

 Since additional constraints can only increase the minimum of a functional,
 the GCM mass must produce smaller action, and thus smaller decay exponent,
 than that of instanton. Any other action obtained with additional
 constraints will also produce larger decay exponent.
  As the ATDHF respects the velocity-momentum
  conditions to the same order to which it is exact, it will produce
  larger $S$ than GCM. 
 The results of calculations seem to support
  this, see e.g. \cite{G90,BSDN07}.
 On the other hand, it is known that by introducing velocities (or momenta)
 as additional generating coordinates, one can show the equivalence of
  such a more general GCM and the ATDHF \cite{BV78,RG79}.


 \section{Inclusion of pairing in the instanton method}

  It is well known that pairing interaction should be taken into account
 if realistic estimates for fission probabilities are to be found.
 In fact, it is pairing that gives the main contribution
 to the mass parameters, as it couples s.p. levels of different
 symmetries when they cross at the Fermi level. At the same time, it
 produces the gap in the quasi-particle spectrum which makes the
 collective motion more adiabatic.
 The proper self-consistent formalism to include pairing in the instanton
 approach is the HFB theory, in which the Slater determinants are replaced by
 the quasi-particle vacua, the many-particle states of undetermined particle
 number, annihilated by a set of operators:
 \begin{equation}
  \alpha_i = \sum_{\mu} ( A^*_{\mu i} a_{\mu} + B^*_{\mu i} a^{+}_{\mu}) ,
 \end{equation}
 where operators $a^+_{\mu}$ refer to some fixed s.p. basis.
  We give here elements of the instanton method for systems with pairing. 
  These include the imaginary-time version of the TDHFB equations, 
  the counterpart of the formula Eq.(\ref{Sinv2})
  for action in terms of familiar HFB states, equations
  in coordinate-momentum variables (analogous to Eqs. (\ref{for2})) 
  and the formulation in terms of a time-even generalized density matrix and 
  a time-odd hermitean opoerator that leads naturally to the adiabatic limit.

 For our purpose it is helpful to notice that the above customary definition
 implies that the HFB vacuum
 $\mid\Psi\rangle \sim \exp(\frac{1}{2}\sum_{\mu \nu}Z_{\mu \nu} 
 a^+_{\mu}a^+_{\nu})\mid 0\rangle$,
 with $Z=B^*A^{*-1}$, depends on matrices $A^*$ and $B^*$,
 while $\langle\Psi\mid$, the corresponding bra, depends on $A$ and $B$.

 \subsection{Imaginary-time TDHFB equations}
 The TDHFB theory is built on the condition of unitarity
 of the time-dependent Bogolyubov transformation and the
 variational principle. The HFB transformation for imaginary time,
 $t\rightarrow -i\tau$, becomes:
  \begin{equation}
  \label{HFB}
  \left(
  \begin{array}{c}
   \alpha^+(\tau) \\
   \alpha(-\tau)  \\
  \end{array}\right) =
  \left(
  \begin{array}{cc}
   A^T(\tau) , &  B^T(\tau) \\
   B^+(-\tau) ,&  A^+(-\tau) \\
  \end{array} \right)
  \left(
  \begin{array}{c}
   a^+ \\
   a  \\
  \end{array}
  \right) ,
 \end{equation}
 where $A(t)$ and $B(t)$ became functions of $\tau$, while their complex
 conjugate $A^*(t)$ and $B^*(t)$ became functions of $-\tau$.
  The unitarity of the HFB transformation in the real-time formalism
 translates to the following condition in the imaginary-time version:
  \begin{equation}
  \label{mat}
  \left(
  \begin{array}{cc}
   A^T(\tau) ,&  B^T(\tau) \\
   B^+(-\tau) ,&  A^+(-\tau) \\
  \end{array}
  \right)^{-1} =
  \left(
  \begin{array}{cc}
   A^*(-\tau) ,&   B(\tau) \\
   B^*(-\tau) ,&  A(\tau) \\
  \end{array}
  \right) .
 \end{equation}
  This equation means that fermionic anticommutation relations
 for operators $a^+_{\mu}, a_{\nu}$ transfer to:
 $\{\alpha_i(-\tau),\alpha_j(-\tau)\}=\{\alpha^+_i(\tau),\alpha^+_j(\tau)\}=0$,
and $\{\alpha_i(-\tau),\alpha^+_j(\tau)\}=\delta_{i j}$ (and {\it vice versa}).
 Denoting ${\cal N}(\tau)$ the imaginary-time HFB transformation
 Eq.(\ref{HFB}), its properties may be concisely written as ${\cal
 N}^{-1}(\tau)={\cal N}^+(-\tau)=\sigma_x {\cal N}^T(\tau)\sigma_x$,
 using the Pauli matrix notation for the block matrix.
 Written as separate conditions these are eight matrix equations which reduce
 to four independent relations in which $\tau$ may be both positive or negative:
 \begin{eqnarray}
 \label{acomm}
  A^+(-\tau)A(\tau)+B^+(-\tau)B(\tau) &  =  & I , \\
\nonumber
  A^T(\tau)B(\tau)+B^T(\tau)A(\tau) &  =  & 0 , \\
\nonumber
  A^*(-\tau)A^T(\tau)+B(\tau)B^+(-\tau) &  =  & I , \\
\nonumber
  A^*(\tau)B^T(-\tau)+B(-\tau)A^+(\tau) &  =  & 0 . \\
\nonumber
 \end{eqnarray}
 The first of those differs from the usual HFB condition as it forces
  anticommutation between annihilation and creation operators of two
 different sets of $\tau$ and $-\tau$. This means that the usual
 relations: $\{\alpha_i(\tau),\alpha^+_j(\tau)\}=\delta_{i j}$ are not ensured.
 However, as shown below, new operators related to $\alpha(\pm\tau)$ may be
 defined, fulfilling usual conditions.

 The variational principle that gives TDHFB equations,
 transformed to imaginary time $t\rightarrow -i\tau$, becomes:
 $\delta\int d\tau\langle\Phi(\tau)\mid\hbar\partial/\partial\tau+{\hat H}\mid
 \Phi(-\tau)\rangle=0$. Calculating variations
 $\delta/\delta A^*_{\mu i}(-\tau)$
 and $\delta/\delta B^*_{\mu i}(-\tau)$ one has to use,
 as in the real-time case, the transformation conditions
 Eqs.(\ref{acomm}) and account for the resulting redundancy of the variables 
 $A$ and $B$. The term with the time derivative that defines action becomes:
 \begin{eqnarray}
  \label{SHFB}
 S/\hbar= \int d\tau \langle \Phi(\tau)\mid \partial_{\tau} \Phi(-\tau)\rangle
 &=&\frac{1}{2}\int d\tau Tr[\partial_{\tau} A^+(-\tau) A(\tau) + 
 \partial_{\tau} B^+(-\tau) B(\tau)]  \\
\nonumber
 &= & -\frac{1}{2}\int d\tau Tr[ A^+(-\tau) \partial_{\tau} A(\tau) + B^+(-\tau) \partial_{\tau}B(\tau)]  .
\nonumber
 \end{eqnarray}
  The matrix element of Hamiltonian $\langle\Phi(\tau)\mid {\hat H}\mid\Phi(-\tau)
\rangle$ is expressed by contractions:
 \begin{eqnarray}
  \langle\Phi(\tau)\mid a^+_{\nu} a_{\mu}\mid\Phi(-\tau)\rangle
  &=\rho_{\mu \nu}(\tau)=& (B^*(-\tau) B^T(\tau))_{\mu \nu}   , \\
\nonumber
  \langle\Phi(\tau)\mid a_{\nu} a_{\mu}\mid\Phi(-\tau)\rangle
  &=\kappa_{\mu \nu}(\tau)=& (B^*(-\tau) A^T(\tau))_{\mu \nu}  , \\
\nonumber
  \langle\Phi(\tau)\mid a^+_{\nu} a^+_{\mu}\mid\Phi(-\tau)\rangle
  &={\tilde \kappa}_{\mu \nu}(\tau)=& (A^*(-\tau) B^T(\tau))_{\mu \nu}  , \\
\nonumber
 \end{eqnarray}
  which, due to conditions (\ref{acomm}), have the following properties when
 regarded as matrices:
 \begin{eqnarray}
 \label{contr}
  \rho(-\tau)&=& \rho^+(\tau) ,  \\
\nonumber
  \kappa^T(\tau)&=& -\kappa(\tau) ,  \\
\nonumber
  {\tilde \kappa}(\tau)&=& \kappa^+(-\tau) .  \\
\nonumber
 \end{eqnarray}
 Using those and proceeding as in the case of TDHFB we arrive at imaginary-time
 TDHFB equations written symbolically (where only the second index of the
 amplitudes is explicit):
  \begin{equation}
  \label{TDHFBi}
  \hbar \partial_{\tau}\left(
  \begin{array}{c}
   A_k(\tau) \\
   B_k(\tau)  \\
  \end{array}\right) +
  \left(
  \begin{array}{cc}
   {\hat t}+{\hat \Gamma}(\tau)  ,&  {\hat \Delta}(\tau)  \\
   -{\hat \Delta}^*(-\tau) ,&  -({\hat t}+{\hat \Gamma}(-\tau))^* \\
  \end{array} \right)
  \left(
  \begin{array}{c}
   A_k(\tau) \\
   B_k(\tau)  \\
  \end{array}
  \right) = E_k
  \left(
  \begin{array}{c}
   A_k(\tau) \\
   B_k(\tau)  \\
  \end{array}
  \right)
 \end{equation}
 where, for a given Hamiltonian, the self-consistent potential:
 $\Gamma_{\mu \nu}(\tau)= \sum_{\gamma \delta}(v_{\mu \gamma \nu \delta}
 -v_{\mu \gamma \delta \nu})\rho_{\delta \gamma}(\tau)$
 and the pairing potential:
$\Delta_{\mu \nu}(\tau)=\sum_{\gamma \delta} 
 v_{\mu \nu \gamma \delta}
\kappa_{\gamma \delta}(\tau)$ have the properties: ${\hat \Gamma}(-\tau)=
 {\hat \Gamma}^+(\tau)$, and ${\hat \Delta}^T(\tau)=-{\hat \Delta}(\tau)$.
 The same properties hold for the mean fields with additional rearrangement terms
 that follow from a density functional.
 These ensure the property ${\hat h}(-\tau)={\hat h}^+(\tau)$ of the
 mean-field Hamiltonian ${\hat h}(\tau)=
  {\hat t}+{\hat \Gamma}(\tau)$, and the same property,
 ${\hat {\bf h}}(-\tau)={\hat {\bf h}}^+(\tau)$ of the total HFB
 mean-field Hamiltonian ${\hat {\bf h}}(\tau)$ given by the matrix in
 Eqs.(\ref{TDHFBi}).
 As a result of this, the equations (\ref{TDHFBi})
 conserve both energy and all relations (\ref{acomm}). The terms
 with constants $E_k$ on the r.h.s. fix the periodicity of solutions and
  these constants are equal to the quasi-particle energies at the metastable
 HFB ground-state.
 The bounce solution to Eqs.(\ref{TDHFBi}) has to be periodic and provide
 a path connecting the HFB ground state $\mid \Psi_{gs}\rangle$ with some HFB
 state $\mid \Phi(\tau=0)\rangle$ at the same energy beyond the barrier.

 \subsection{Variational principle}
 In a similar way as in the HF case, one can deduce the minimum principle for
  action under conditions
 of constant energy and fulfilled Eqs.(\ref{TDHFBi}) for $0<\tau<T/2$.
 The redundancy of variables $A,B$ complicates the Hamilton equations, but 
 the following relations hold: $-2\delta{\cal H}=\sum_k
 (\langle\delta{\cal W}_k(-\tau)\mid{\hat {\bf h}}(\tau)\mid{\cal W}_k(\tau)\rangle
 +\langle{\cal W}_k(-\tau)\mid{\hat {\bf h}}(\tau)\mid\delta{\cal W}_k(\tau)\rangle)$ 
 and
  $-2\delta S=\hbar(\sum_k
 (\langle\delta{\cal W}_k(-\tau)\mid\partial_{\tau}{\cal W}_k(\tau)\rangle
 -\langle\partial_{\tau}[{\cal W}_k(-\tau)]\mid\delta{\cal W}_k(\tau)\rangle)$,
 with ${\cal W}_k$ denoting the vector composed of $(A_k,B_k)$. Since taking a 
 formal variation of $S+{\cal H}$ with respect to $\delta {\cal W}^*_k$ and 
 $\delta {\cal W}_k$ leads to the correct equations (\ref{TDHFBi}), the 
 arguments of sect. IV can be repeated and one obtains the same constraints 
 that specify bounce as the minimum of action (note  
 $\langle{\cal W}_k(-\tau)\mid{\cal W}_l(\tau)\rangle=\delta_{k l}$).

 The first of Eqs.(\ref{acomm}) means that
 $\langle\Phi(\tau)\mid\Phi(-\tau)\rangle=1$.
 Since these two HFB states are different, the imaginary-time HFB
 transformation determined by the matrices $A(\pm\tau)$ and $B(\pm\tau)$
 cannot be unitary. However, it may be related to a normal unitary
 HFB transformation given by some matrices $U(\tau),V(\tau)$ via some
 invertible, though non-unitary matrices $C(\tau)$. Let us suppose a relation:
 \begin{equation}
  \alpha^+_i(\tau)=\sum_j C_{j i}(\tau) \beta^+_j(\tau) ,
 \end{equation}
  with quasi-particle creation operators $\beta^+_i(\tau)$ related via some
  $U(\tau)$ and $V(\tau)$ matrices to $a^+_{\mu},a_{\mu}$, namely 
 [cf Eq.(\ref{HFB})]:
  \begin{equation}
  \label{HFB1}
  \left(
  \begin{array}{c}
   \alpha^+(\tau) \\
   \alpha(-\tau)  \\
  \end{array}\right) =
  \left(
  \begin{array}{cc}
   (U(\tau)C(\tau))^T ,&  (V(\tau)C(\tau))^T \\
   (V(-\tau)C(-\tau))^+ ,&  (U(-\tau)C(-\tau))^+ \\
  \end{array} \right)
  \left(
  \begin{array}{c}
   a^+ \\
   a  \\
  \end{array}
  \right) .
 \end{equation}
  It follows that $U(\tau), V(\tau)$ define the same $Z(\tau)$ as $A(\tau)$ 
 and $B(\tau)$ do and that $U(\tau)^+U(\tau)+V(\tau)^+V(\tau)=
 C^{+-1}(\tau)(A^+(\tau)A(\tau)+B^+(\tau)B(\tau))C^{-1}(\tau)$. If one chooses 
  $C(\tau)$ that transforms the hermitean matrix 
 $A^+(\tau)A(\tau)+B^+(\tau)B(\tau)$ 
 to the unit matrix, then $U(\tau)$ and $V(\tau)$ become matrices of a 
  standard HFB transformation. Now, the first of Eqs.(\ref{acomm}) means that:
  \begin{equation}
   \label{mat1}
   (U(-\tau)^+U(\tau)+V(-\tau)^+V(\tau))^{-1} = C(\tau)C(-\tau)^+ ,
  \end{equation}
  while three other follow from this and from the HFB
 properties of matrices $U(\tau),V(\tau)$ and $U(-\tau),V(-\tau)$.
 The second equation (\ref{acomm}) is just the condition of the antisymmetry 
 of $Z(\tau)$, the equations three and four:
 $(I+Z^+(\tau)Z(-\tau))^{-1}+Z^+(\tau)(I+Z(-\tau)Z^+(\tau))^{-1}
 Z(-\tau)=I$
 and the antisymmetry of matrices:
 $Z(\tau)^+(I+Z(-\tau)Z^+(\tau))^{-1}$ and 
 $(I+Z(-\tau)Z^+(\tau))^{-1}Z(-\tau)$, follow from the previous two.

 Using the same reasoning as the one leading to Eq.(\ref{Sinv2}), 
 instanton action
 (\ref{SHFB}) can be expressed in terms of the normalized HFB states
 $\mid\Psi(\tau)\rangle$, defined by $U(\tau)$ and $V(\tau)$, using relation
  (\ref{mat1}):
 \begin{equation}
 \label{SHFB1}
  S/\hbar=-\frac{1}{2}\Re\int_{-T/2}^{T/2} d\tau Tr[(U^+(-\tau)U(\tau)+V^+(-\tau)V(\tau))^{-1}
(U^+(-\tau) \partial_{\tau}U(\tau) + V^+(-\tau)\partial_{\tau}V(\tau))] ,
 \end{equation}
 where we have omitted the integral of $\partial_{\tau}\ln\det C(\tau)$ between
  the endpoints, as it is purely imaginary.

 The contractions Eq.(\ref{contr}) can be expressed through
$U(\pm\tau),V(\pm\tau)$ and the corresponding HFB states $\Psi(\pm\tau)$ in
 the following way:
 \begin{eqnarray}
\rho_{\mu \nu}&=(V^*(-\tau)({\tilde U}(\tau)^T)^{-1}V^T(\tau))_{\mu \nu}
 =&\frac{\langle\Psi(\tau)\mid a^+_{\nu} a_{\mu}\mid\Psi(-\tau)\rangle}
  {\langle\Psi(\tau)\mid\Psi(-\tau)\rangle}  , \\
\nonumber
\kappa_{\mu \nu}&=(V^*(-\tau)({\tilde U}(\tau)^T)^{-1}U^T(\tau))_{\mu \nu}=&
  \frac{\langle\Psi(\tau)\mid a_{\nu} a_{\mu}\mid\Psi(-\tau)\rangle}
  {\langle\Psi(\tau)\mid\Psi(-\tau)\rangle}  , \\
\nonumber
{\tilde \kappa}_{\mu \nu}&=
(U^*(-\tau)({\tilde U}(\tau)^T)^{-1}V^T(\tau))_{\mu \nu}=&
  \frac{\langle\Psi(\tau)\mid a^+_{\nu} a^+_{\mu}\mid\Psi(-\tau)\rangle}
  {\langle\Psi(\tau)\mid\Psi(-\tau)\rangle}  , \\
\nonumber
 \end{eqnarray}
 where the matrix ${\tilde U}(\tau)=U^+(-\tau)U(\tau)+V^+(-\tau)V(\tau)$
 is related to the overlap of standard HFB states via:
 $\langle\Psi(\tau)\mid\Psi(-\tau)\rangle=[\det{\tilde U}(\tau)]^{1/2}$
 \cite{RingS}.

 Now, one can treat (\ref{SHFB1}) as a functional on trial fission paths
 $\Psi(\tau)$, defined by two families of HFB states $\Psi_1(\tau)$ and
 $\Psi_2(\tau)$ for $0<\tau<T/2$
  \begin{equation}
   \Psi(\tau) = \left\{ \begin{array}{cc}
            \Psi_{1}(-\tau) & \mbox{for} \; \tau<0 , \\
            \Psi_{2}(\tau)   & \mbox{for} \; \tau>0 , \\
                   \end{array} \right\}
  \end{equation}
 smoothly connecting some HFB state $\Phi(0)$ beyond the barrier
 at energy $E_{gs}$ to the metastable ground state $\Psi_{gs}$,
 and fulfilling the condition of constant energy overlap and the 
 Eqs.(\ref{TDHFBi}) for $\Psi_{2}(\tau)$.
  Taking ${\tilde U}(\tau)=
 U_1^+(\tau)U_2(\tau)+V_1^+(\tau)V_2(\tau)$ for $\tau>0$, and having
 ${\tilde U}(\tau)={\tilde U}^+(-\tau)$ for $\tau<0$, one can calculate
 action as:
 \begin{equation}
 \label{SHFB2}
  S/\hbar=-\Re\int_{0}^{T/2} d\tau Tr[{\tilde U}^{-1}(\tau)
(U_1^+(\tau)\partial_{\tau}U_2(\tau) + V_1^+(\tau)\partial_{\tau}V_2(\tau))] .
 \end{equation}
 The minimization of this action over fission paths that fulfil constraints
 should reproduce the bounce action. Its value for a trial path that
 satisfies constraints is an upper bound for the bounce decay exponent.

 \subsection{Coordinate and momentum variables}
   The coordinate-momentum variables may be introduced in a similar way as in
 section V. Decomposing amplitudes into $\tau$-even
 and $\tau$-odd components,
 $A(\tau)=A_+(\tau)-A_-(\tau)$, $A(-\tau)=A_+(\tau)+A_-(\tau)$,
 $B(\tau)=B_+(\tau)-B_-(\tau)$, $B(-\tau)=B_+(\tau)+B_-(\tau)$, with
   $A_+$ and $B_+$ matching $\Psi_{gs}$ at $\tau=\pm T/2$ and
  $\Phi(0)$ at $\tau=0$, and $A_-=B_-=0$ at $\tau=0,\pm T/2$, one obtains the
  system of equations (with only the second index of the amplitudes made
  explicit)
  \begin{equation}
  \label{TDHFBii}
  \hbar \partial_{\tau}\left(
  \begin{array}{c}
   A_{+ k} \\ B_{+ k} \\ A_{- k} \\ B_{- k} \\
  \end{array}\right) =
  \left(
  \begin{array}{cccc}
   -{\hat h}_A  ,&  -{\hat \Delta}_- , & {\hat h}_R-E_k , & {\hat \Delta}_+  \\
   -{\hat \Delta}_-^* ,&  -{\hat h}_A^* , & -{\hat \Delta}_+^* , &
   -{\hat h}_R^*-E_k \\
   {\hat h}_R-E_k , & {\hat \Delta}_+ , & -{\hat h}_A , & -{\hat \Delta}_-  \\
   -{\hat \Delta}_+^* ,& -{\hat h}_R^*-E_k , & -{\hat \Delta}_-^* , &
   -{\hat h}_A^* \\
  \end{array} \right)
  \left(
  \begin{array}{c}
   A_{+ k} \\ B_{+ k} \\ A_{- k} \\ B_{- k} \\
  \end{array}
  \right)  ,
 \end{equation}
  with the mean fields ${\hat h}={\hat h}_R+{\hat h}_A$ and
  ${\hat \Delta}={\hat \Delta}_+ + {\hat \Delta}_-$, with
  ${\hat \Delta}_+(-\tau)={\hat \Delta}_+(\tau)$ and
  ${\hat \Delta}_-(-\tau)=-{\hat \Delta}_-(\tau)$. In a similar way as for  
  Eqs. (\ref{for2}), the first two Eqs. (\ref{TDHFBii}) connect velocities
  $\partial_{\tau} A_{+ k}$, $\partial_{\tau} B_{+ k}$ with momenta
  $A_{- k}$ and $B_{- k}$, showing that they all, together with the $\tau$-odd
  mean-field potentials ${\hat h}_A$ and ${\hat \Delta}_-$, are proportional
  to the collective velocity ${\dot Q}$.
  In the coordinate-momentum representation, these are the constraints
  that must be imposed on trial fission paths to assure that bounce
  provides the minimum of the action functional.
  The Eqs.(\ref{acomm}) written in terms of new amplitudes become eight
  relations which may be combined to four $\tau$-even and four $\tau$-odd
  equations,  e.g. the first Eq.(\ref{acomm}) leads to
  $A_+^+A_+-A_-^+A_-+B_+^+B_+-B_-^+B_-=I$ and
  $A_-^+A_+-A_+^+A_-+B_-^+B_+-B_+^+B_-=0$, etc.

   Let us call the diagonal and off-diagonal submatrices of the matrix in
   Eq. (\ref{TDHFBii}) $-{\hat {\bf h}}_A$ and ${\hat {\bf h}}_R$. From
   symmetries and definitions it is clear that
    ${\hat {\bf h}}_R(\tau)$ is hermitean and time-even and
    ${\hat {\bf h}}_A(\tau)$ - antihermitean and time-odd.
 In imaginary-time TDHFB, the operator ${\hat {\bf h}}_A$ is the 
 generalization of the Thouless-Valatin mean field ${\hat h}_A$ of the ATDHF.

  Denote the vector built of $A_{+ k}$ and $B_{+ k}$ as $\Theta_k$ and the
  one built of $A_{- k}$ and $B_{- k}$ as $\Xi_k$, i.e. ${\cal W}_k(\tau)=
   \Theta_k(\tau)-\Xi_k(\tau)$. Then the Eqs.(\ref{TDHFBii})
  take the form
   \begin{eqnarray}
   \label{TDHFBiii}
 \hbar\partial_{\tau}\Theta_k & = & -{\hat {\bf h}}_A \Theta_k +
  ({\hat {\bf h}}_R-E_k) \Xi_k , \\ \nonumber
 \hbar\partial_{\tau}\Xi_k & = & ({\hat {\bf h}}_R-E_k) \Theta_k -
  {\hat {\bf h}}_A \Xi_k . \\ \nonumber
   \end{eqnarray}
  The variation of energy written in terms of $\Theta_k$ and 
   $\Xi_k$ reads
 \begin{equation}
 2\delta {\cal H} = \sum_k\left( \langle\delta\Theta_k\mid {\hat {\bf h}}_A\Xi_k
\rangle-\langle\delta\Theta_k\mid {\hat {\bf h}}_R\Theta_k\rangle
      -\langle\delta\Xi_k\mid {\hat {\bf h}}_A\Theta_k\rangle +
  \langle\delta\Xi_k\mid {\hat {\bf h}}_R\Xi_k\rangle + c.c.\right).
 \end{equation}
  The three last terms, together with their complex conjugate, 
  contribute at the second order in $\tau$-odd components, assuming 
   $\Xi_k$ and $\delta \Xi_k$ being of the first, and $\delta \Theta_k$ of the 
  second order of smallness.
  Owing to the $\tau$-parity of the amplitudes, and after integrating by parts,
  action reads:
 \begin{equation}
  S/\hbar=-2\Re \int_{0}^{T/2} d\tau Tr[ A_-^+(\tau) \partial_{\tau} A_+(\tau)
  + B_-^+(\tau) \partial_{\tau}B_+(\tau)]  .
  \end{equation}
  This can be expressed as
  $S=-\hbar\Re\int_{-T/2}^{T/2} d\tau \sum_k \langle\Xi_k\mid\partial_{\tau}
 \Theta_k\rangle$, i.e. it is imaginary-time TDHFB action in the form
  $\int p_i dq_i$. Substituting $\Xi_k$ from the first Eq.(\ref{TDHFBiii}) 
  one can obtain the cranking-like expression for action as in sect. V.

 \subsection{Adiabtic expansion and limit}
  The above formulas are a copy of those in sections V and VI, up to the 
  common factor
  $(-1/2)$ appearing in the expressions for $S$ and $\delta {\cal H}$. 
  Hence, after showing that the operator that maps amplitudes at $\tau$ onto 
 those at $-\tau$ is hermitean one could represent HFB bounce in terms of 
  $\tau$-even amplitudes and
    a $\tau$-odd hermitean operator ${\hat S}$, as in sect. VI, and repeat the
  whole reasoning on the adiabatic limit of the instanton method.
  (To emphasize the analogy, we keep the same notation for the time-odd 
  operator as in HF, although it acts in the enlarged space.)
   
   The argument goes as follows: The HFB transformation
  from operators $(\alpha^+(\tau),\alpha(-\tau))$ to
  $(\alpha^+(-\tau),\alpha(\tau))$ is ${\cal N}(-\tau){\cal
  N}^{-1}(\tau)={\cal N}(-\tau){\cal N}^+(-\tau)$ (cf Eq. (\ref{mat})), 
  indeed hermitean.
  Calling this transformation $\exp(2{\cal S}(\tau))$, with ${\cal S}(\tau)$ hermitean, and considering its
  inverse, we have ${\cal S}(-\tau)=-{\cal S}(\tau)$. Then, we find that $\exp({\cal
   S}(\tau)){\cal N}(\tau)=\exp({\cal S}(-\tau)){\cal N}(-\tau)$, so
  calling this $\tau$-even transformation ${\bar {\cal N}}(\tau)$, we
  have ${\bar {\cal N}}^{-1}(\tau)={\bar {\cal N}}^+(\tau)$, so ${\bar {\cal N}}(\tau)$
  is a regular HFB transformation. Denoting its amplitudes $u$ and $v$,
    we have
\begin{equation}
 \left(
 \begin{array}{cc}
  A^T(\tau) ,& B^T(\tau) \\
  B^+(-\tau) ,& A^+(-\tau) \\
 \end{array}\right) = \exp(-{\cal S}(\tau))
 \left(
 \begin{array}{cc}
  u^T(\tau) ,& v^T(\tau) \\
  v^+(\tau) ,& u^+(\tau) \\
   \end{array} \right) .
\end{equation}
The properties of ${\cal N}(\tau)$ and ${\bar {\cal N}}(\tau)$ imply
$\sigma_x {\cal S}^T(\tau) \sigma_x = -{\cal S}(\tau)$.
 As we need a relation between amplitudes and these form columns of
the matrices ${\cal N}^T(\tau)$ and ${\bar {\cal N}}^T(\tau)$, we
notice that ${\cal N}^T(\tau)=[{\bar {\cal N}}^T(\tau)\exp(-{\cal
S}^T(\tau))({\bar {\cal N}}^T)^{-1}(\tau)]{\bar {\cal N}}^T(\tau)$,
and that the matrix ${\bar {\cal N}}^T(\tau)\exp(-{\cal
S}^T)(\tau))({\bar {\cal N}}^T)^{-1}(\tau)$ is hermitean owing to
the HFB property of ${\bar {\cal N}}(\tau)$. Moreover, due to this
property, one has ${\cal N}^T(\tau)=\exp(-{\hat S}(\tau)){\bar {\cal
N}}^T(\tau)$ with the hermitean, $\tau$-odd
 ${\hat S}(\tau)={\bar {\cal N}}^T(\tau){\cal S}^T(\tau) {\bar {\cal
 N}}^*(\tau)$. It follows from the properties of ${\cal N}$ and ${\cal S}$ that
  $\sigma_x {\hat S}^T(\tau) \sigma_x = -{\hat S}(\tau)$. Thus 
 \begin{equation}
  {\hat S}=\left(\begin{array}{cc} {\hat s} & {\hat r} \\  
                                  -{\hat r}^* & -{\hat s}^* \\
  \end{array} \right) ,
 \end{equation}
 with ${\hat s}^+={\hat s}$, and ${\hat r}^T=-{\hat r}$.
 With this ${\hat S}(\tau)$, we have the expected relations
\begin{equation}
 \left(
 \begin{array}{c}
  A_k(-\tau) \\
  B_k(-\tau)  \\
 \end{array}\right) = \exp({\hat S}(\tau))
 \left(
 \begin{array}{c}
  u_k(\tau)  \\
  v_k(\tau) \\
 \end{array} \right)  ;
 \left(
 \begin{array}{c}
  A_k(\tau) \\
  B_k(\tau)  \\
 \end{array}\right) = \exp(-{\hat S}(\tau))
 \left(
 \begin{array}{c}
  u_k(\tau) \\
  v_k(\tau)  \\
 \end{array}
 \right) ,
\end{equation}
 where only the second index of the amplitudes is shown. With these, all the 
 results of the section VI can be repeated for imaginary-time TDHFB. 
 In particular, the integrand of the action integral $S$,   
 which in terms of the amplitudes ${\cal W}_{0 k}=(u_k,v_k)$ and the operator 
 ${\hat S}$ reads  
 $-\frac{\hbar}{2}\sum_k(\langle \partial_{\tau}{\cal W}_{0 k}\mid{\hat S}
 \mid{\cal W}_{0 k} \rangle+ c.c.)$ is equal to 
 $-2({\cal H}-{\cal H}_0)$ at the second order in ${\hat S}$, hence positive. 
 The Eqs. (\ref{TDHFBi}) take exactly the form (\ref{eqsymS}) of the 
 imaginary-time TDHF, with obvious replacements of ${\cal W}_{0 k}$ for 
 $\psi_{0 k}$ and ${\hat {\bf h}}$ for ${\hat h}$. 
 They reduce to the form (\ref{eqS}) at the second order in ${\hat S}$. 

 The TDHFB equations may be also formulated in terms of the generalized density 
 matrix. The counterpart of the HFB density matrix in the imaginary-time 
 formalism is
\begin{equation}
 \left(
 \begin{array}{cc}
  \rho(\tau), & \kappa(\tau) \\
  -\kappa^*(-\tau), & I-\rho^*(-\tau)  \\
 \end{array}\right) =\left(
 \begin{array}{c}
  B^*(-\tau) \\
  A^*(-\tau)  \\
 \end{array}\right) ( B^T(\tau) , A^T(\tau) ) = (\sigma_x\exp({\hat
 S}^*(\tau))\sigma_x){\cal R}_0(\tau)(\sigma_x\exp(-{\hat
 S}^T(\tau))\sigma_x)  ,
\end{equation}
 with ${\cal R}_0(\tau)$ the HFB density matrix corresponding to 
 ${\bar {\cal N}}(\tau)$.
 Owing to the property of ${\hat S}$, it is equal to 
 $e^{-{\hat S}(\tau)}{\cal R}_0(\tau)e^{{\hat S}(\tau)}$.
 This non-hermitean quantity, call it ${\widetilde {\cal R}}$, apart from not 
 being any HFB density matrix, is an analogue 
 (note that ${\widetilde {\cal R}}^2={\widetilde {\cal R}}$) 
 of the density matrix in the ATDHFB theory \cite{B65}, 
 ${\cal R}=e^{i{\hat \chi}}{\cal R}_0 e^{-i{\hat \chi}}$. In terms of it,
 Eqs.(\ref{TDHFBi}) read:
 $\hbar\partial_{\tau} {\widetilde {\cal R}}+[{\hat {\bf h}},{\widetilde {\cal
 R}}]=0$. The $\tau$-odd part of this equation, linear in ${\hat S}$, 
 obtained by expanding ${\widetilde {\cal R}}={\cal R}_0-
 [{\hat S},{\cal R}_0]+...$ and discarding the second order quantity 
 $[{\hat {\bf h}}_0,{\cal R}_0]$,  
\begin{equation}
 \hbar\partial_{\tau} {\cal R}_0+[[{\hat S},{\hat {\bf h}}_0]+ 
 {\hat {\bf h}}_A,{\cal R}_0]=0 , 
 \end{equation}
 is an alternative form of the second Eq.(\ref{eqS}) in terms of  
 ${\cal R}_0$ and ${\hat S}$.  
 Its solution is identical to the ATDHFB solution,
 ${\hat S}={\hat \chi}$. 
 This follows directly from the structure of the building blocks of 
 the Thouless-Valatin mean field ${\hat {\bf h}}_A$. One has  
 ${\hat h}_A=Tr({\tilde v}\rho_1)$ and $\Delta_{-\alpha \beta}=
 \sum_{\gamma\delta} v_{\alpha \beta \gamma \delta}\kappa_{1 \gamma \delta}$,
  with $\rho_1=-[{\hat s},\rho_0]+{\hat r}\kappa_0^*-\kappa_0{\hat r}^*$, 
 $\kappa_1=\rho_0{\hat r}+{\hat r}(\rho_0^*-1)-{\hat s}\kappa_0-
 \kappa_0{\hat s}^*$.
 Since, in ATDHFB, ${\cal R}_1=i[{\hat \chi},{\cal R}_0]$, 
  one has ${\hat {\bf h}}_A=i{\hat {\bf h}}_1$, where 
 ${\hat {\bf h}}_1$ is the ATDHFB time-odd mean field for  
 ${\hat \chi}={\hat S}$. Thus, the adiabatic TDHFB instanton 
 method produces mass given by:   
  $mass\times{\dot Q}^2=\frac{\hbar}{2}Tr({\dot {\cal R}}_0{\hat S})$,  
  equal to the ATDHFB mass, cf \cite{DS}. In the zero pairing limit     
  this mass reduces to the ATDHF value 
 $\hbar Tr({\dot \rho}_0{\hat s})/{\dot Q}^2$.  

  A reasoning similar to the one presented in section VII shows that, 
  within the GCM approach, a use of some $\tau$-even pairing variable 
 (for example, the pairing gap) as a generator coordinate, without fulfilling 
  the velocity-momentum relations, will
 lead to a smaller decay exponent than that for bounce.

 \section{Conclusions}

  We have presented the instanton method for nuclear fission in various
  representations.  This has allowed for some comparisons with other
  methods commonly used in fission studies. We have also sketched the
  imaginary-time version of the TDHFB theory, which allows to include pairing.

  There are many similarities between the instantons describing quantum
  tunneling and the periodic TDHF solutions.
  Both appear as a result of the quasi-classical approximation, find a
  natural formulation in terms of time-even coordinates and
  time-odd momenta and reduce to the same time-odd ATDHF equation in the
  lowest order in momenta. The ATDHF equation for a path, which should be
  fulfilled for consistency, is usually not checked for static
  paths constructed by means
  of the CHF. When the velocity-momentum equations require small momenta
   that violate energy conservation, this means that the
  chosen path is far from instanton.

 The main difference between the two methods is that in
 quantum tunneling there is no single HF state or density
 matrix, but one deals with two different states, bra and ket.
 This happens to be the very reason for the existence of the minimum
 principle: it defines the minimal driving of one state by the other, necessary
 for tunneling. 
  Instanton action turns out to be a minimum of the action functional
  when the constraints of constant energy and velocity-momentum
  relations are imposed on trial fission paths.
  Action calculated for any such path would provide an upper bound
  for the decay exponent. We argue that the ATDHF (ATDHFB) mass
  respects those constraints, while the GCM+GOA mass does not.
   The main practical problem is how to construct trial paths fulfilling
   the constraints.

  The need for two Slater determinants for instanton leads to another
  important difference between the mean-field studies of oscillations and
  quantum tunneling:
 The instanton method relies on the off-diagonal matrix elements of the
 Hamiltonian, which are beyond the usual scope of the mean-field
 theory. To use instantons in practice, one has to define various off-diagonal
 matrix elements of the commonly used effective interactions, like,
 for example, of the density-dependent term of the Skyrme-like force
 (for its possible definitions see \cite{DB}) and of the Coulomb-exchange 
 interaction.

  When comparing the instanton method to theories of large amplitude collective
 motion (LACM) one has to recognize that the aims of the latter are much
 wider than those of the former \cite{RG78,RG79}.
 In LACM, equations for the collective path or action are a source
 of formulas for potential and inertia tensor of an effective
 Hamiltonian in a restricted set of deformation coordinates and conjugate
 momenta. Often the next step consists in the
 requantization. The supposed universality of the so conceived effective
 theory for LACM underlies the whole procedure.
 On the contrary, instanton should be found once for a studied decay.
 No interpretation of the integrand in the action formula as
 $mass\times{\dot Q}^2$ is necessary.
 It could be even dangerous, as in some representations of instanton
 these integrands are piecewise negative. Only the value of the integral
 has the physical significance and this does not depend on the representation.

 Of course, one could extract collective inertia from action represented with
 a positive integrand, but the positivity is obvious only in the
 adiabatic limit. In a general case, action Eq.(\ref{action}) contains
 momenta $\xi_k$ to all even orders, and the higher order terms become
 naturally more important for higher barriers.
 Hence one expects that mass also depends on the barrier height, or energy,
 when tunneling from excited states is considered.
 A small energy dependence of mass is seen even for
 highly collective Bose-Einstein condensate \cite{JS}.

  For pairing gaps of $\sim$ 1 MeV and for not too high fission barriers, the
  adiabatic approximation may be satisfactory for many fission paths.
  Then it may appear that the most important in the search for
  instanton is the exploration of a sufficiently rich family of paths,
  preferably with as few preserved symmetries as possible, while
   ATDHFB action (including Thouless-Valatin terms) is a sufficient estimate
  of the instanton action.

  Even if this is true, fission of odd-$Z$ or odd-$N$ nuclei will require
  much more effort to understand, within the instanton method, a dramatic
  significance of the odd fermion and of the specific mean fields induced by 
  it that break time-reversal invariance.

  It is clear that the method considered here is applicable
  to quantum tunneling in any fermion system, provided it has a meaningful
  mean-field description.
   Extensions to include thermal effects and decay form excited states
   seem also straightforward. The real progress of the method will depend
  on practical solutions.

 {\bf Acknowledgments:}

 The author is grateful to J. Dobaczewski for interesting discussions.
 This work was supported in part by the Polish Committee for Scientific
 Reserch (KBN) Grant No. 1P03B06427 and the Polish Ministry of Science,
  by the National Nuclear Security
 Administration under the Stewardship Science Academic Alliances
 program through the U.S. Department of Energy Research Grant
 DE-FG03-03NA00083; by the U.S. Department of Energy under Contract
 No. DE-AC05-00OR22725 with UT-Battelle, LLC (Oak Ridge National
 Laboratory).


\end{document}